\documentclass[a4paper,11pt]{article}
\usepackage{jheppub} 
\usepackage[T1]{fontenc} 
\usepackage{mathtools,amsmath,amssymb,amsfonts,dsfont}
\usepackage{physics}
\usepackage[utf8]{inputenc}
\usepackage[normalem]{ulem}
\usepackage[dvipsnames]{xcolor}
\usepackage{mathrsfs}
\usepackage{footnote}
\usepackage{fontenc}
\usepackage{tabularx}
\usepackage{lscape}
\usepackage{tikz}
\usepackage[makeroom]{cancel}
\usepackage[marginal,stable,bottom]{footmisc}

\newcommand{\X}{\mathcal{X}}
\newcommand{\Lm}{\mathcal{L}}
\newcommand{\XH}{X_{\textrm{\tiny H}}}
\newcommand{\XP}{X_{\textrm{\tiny P}}}

\DeclareMathOperator{\extdm}{d}
\newcommand{\extd}{\extdm \!}

\definecolor{rosy}{RGB}{230,235,252}
\definecolor{myframetitle}{RGB}{90,89,170}
\definecolor{myblocktitle}{RGB}{140,185,249}
\definecolor{mytitle}{RGB}{10,80,26}

\definecolor{darkgreen}{RGB}{27,130,45}
\definecolor{darkblue}{rgb}{0,0,0.3}
\definecolor{darkred}{rgb}{0.7,0,0}

\definecolor{light gray}{RGB}{220,220,220}
\definecolor{dark purple}{RGB}{108,0,217}
\definecolor{pink}{RGB}{190,20,100}
\definecolor{orang}{RGB}{193,63,0}
\definecolor{green}{RGB}{11,98,17}
\definecolor{darkpink}{RGB}{153,0,76}
\definecolor{bluegreen}{RGB}{0,102,102}
\definecolor{greenlagan}{RGB}{0,102,0}
\definecolor{redgreen}{RGB}{102,102,0}
\definecolor{Redgreen}{RGB}{153,76,0}
\definecolor{vividviolet}{rgb}{0.62, 0.0, 1.0}
\definecolor{amaranth}{rgb}{0.9, 0.17, 0.31}
\definecolor{palatinateblue}{rgb}{0.15, 0.23, 0.89}
\definecolor{brightpink}{rgb}{1.0, 0.0, 0.5}
\definecolor{cornflowerblue}{rgb}{0.39, 0.58, 0.93}
\definecolor{deepcarminepink}{rgb}{0.94, 0.19, 0.22}
\definecolor{radicalred}{rgb}{1.0, 0.21, 0.37}

\usepackage[most]{tcolorbox}

\tcbset{highlight math style={left=02mm,right=02mm,top=02mm,bottom=02mm}} 
\hypersetup{ linktoc=all,
    colorlinks, linkcolor={palatinateblue},
    citecolor={brightpink}, urlcolor={amaranth}
}

  \makeatletter  
\gdef\@fpheader{}  
\makeatother 

\newcommand{\eq}[2]{\begin{equation} #1 \label{#2} \end{equation}}

\begin{document}

\title{Carroll dilaton supergravity in two dimensions}

\author[\clubsuit,\diamondsuit]{Daniel Grumiller,}
\author[\clubsuit,\diamondsuit]{Luciano Montecchio,}
\author[\clubsuit,\diamondsuit,\heartsuit]{and Mohaddese Shams Nejati}

\affiliation[\clubsuit]{Institute for Theoretical Physics, TU Wien, Wiedner Hauptstrasse 8, A-1040 Vienna, Austria}
\affiliation[\diamondsuit]{Erwin Schr\"odinger International Institute for Mathematics and Physics, Boltzmanngasse 9, A-1090 Vienna, Austria}
\affiliation[\heartsuit]{Department of Physics, Institute for Advanced Studies in Basic Sciences (IASBS),
P.O.~Box 45137-66731, Zanjan, Iran}

\emailAdd{grumil@hep.itp.tuwien.ac.at,
luciano.montecchio@tuwien.ac.at,
m{\textunderscore}shams@iasbs.ac.ir}

\abstract{
We construct and discuss generic $\mathcal{N} = 1$ and $\mathcal{N} = 2$ Carroll dilaton supergravity in two dimensions. We apply our general results to the supersymmetric Carroll--Jackiw--Teitelboim model, including a discussion of specific boundary conditions. For $\mathcal{N} = 2$ Carroll dilaton supergravity, we find two versions, dubbed ``democratic'' and ``despotic''.
}

\preprint{TUW--24--03}

\maketitle
\section{Introduction}

Supergravity in two dimensions is a venerable subject dating back to the early 1990ies \cite{Park:1993sd}, with precursors in the late 1970ies \cite{Howe:1978ia}, shortly after the discovery of supergravity \cite{Freedman:1976xh,Deser:1976eh}. Generic two-dimensional (2d) dilaton supergravity is efficiently described as a graded Poisson sigma (gPSM) model \cite{Ikeda:1994dr,Izquierdo:1998hg,Strobl:1999zz,Ertl:2000si,Bergamin:2003mh}.

Standard supergravity is based on the supersymmetric version of the Poincar\'e algebra. The singular limit of vanishing speed of light contracts the Poincar\'e algebra to the Carroll algebra \cite{LevyLeblond1965,Gupta1966}. 

Given the current interest in Carrollian physics and holography (see, for instance, the talks at a recent program at the Erwin-Schr\"odinger Institute \cite{ESI2024}), it is natural to construct and discuss Carroll supergravity. The lowest spacetime dimension with a meaningful notion of Carroll supergravity is two, and perhaps the simplest such model is the Carroll--Jackiw--Teitelboim dilaton supergravity model \cite{Ravera:2022buz}. We postpone more concrete motivations for and possible applications of 2d Carroll dilaton supergravity to the conclusions. Furthermore, the three-dimensional case has also been studied by considering the AdS Carroll Chern-Simons supergravity theory \cite{Ravera:2019ize, Ali:2019jjp}.

The main goal of the present work is to construct and discuss generic $\mathcal{N} = 1$ and $\mathcal{N} = 2$ Carroll dilaton supergravity in two dimensions.

This work is organized as follows. In Section \ref{se:2}, we focus on $\mathcal{N}=1$ Carroll dilaton supergravity, starting with Carroll--Jackiw--Teitelboim (CJT) supergravity and then generalizing it to generic Carroll dilaton supergravity by virtue of consistent deformations. We also check compatibility with the Carroll contraction of standard dilaton supergravity, discuss solutions to the equations of motion (EOM), and address selected boundary aspects. In Section \ref{se:3}, we consider $\mathcal{N}=2$ Carroll dilaton supergravity and find two different options, a ``democratic'' one where all spinors are created equal, and a ``despotic'' one where this is not the case. In Section \ref{se:4}, we conclude. Our notations and conventions are summarized in Appendix \ref{Appendix:Notation}.

\section{\texorpdfstring{$\boldsymbol{\mathcal{N}=1}$}{N=1} Carroll dilaton supergravity}\label{se:2}

This Section is devoted to $\mathcal{N}=1$ Carroll dilaton supergravity. It is organized as follows: In Subsection \ref{se:2.1}, we commence with the symmetries of CJT supergravity (sCJT), continue with its BF formulation, and finish with its reinterpretation as a gPSM. In Subsection \ref{se:2.2}, we consistently deform the sCJT gPSM by solving the constraints from (graded) Jacobi identities, focusing on a specific family of models formulated in terms of a dilaton prepotential. We also show that the action for general dilaton supergravity can be understood as a Carrollian limit of Lorentzian general dilaton supergravity, i.e., Carroll contraction and gPSM deformations commute. In Subsection \ref{se:2.3}, we study various aspects of Carroll dilaton supergravity: its first- and second-order formulations, the EOM and their solutions, as well as a list of selected relevant models. In Subsection \ref{se:2.4}, we address an example for boundary conditions in sCJT.

\subsection{Carroll--Jackiw--Teitelboim supergravity}\label{se:2.1}

We start by introducing the Lorentzian $\mathcal{N} = 1$ supersymmetric extension of the AdS$_{2}$ algebra \cite{Bergshoeff:2015wma},
\begin{subequations}
    \label{algebra}
\begin{align}
\big[ K, P_0 \big] &=  P_{1} & \big[ K, P_1 \big] &=  P_{0} & \big[ P_0, P_1 \big] &= \frac{1}{\ell^2} K\\
\big[ K, Q_\alpha \big] &=  \frac{1}{2} (\gamma_\ast Q)_\alpha & \big[ P_{a}, Q_\alpha \big] &= \frac{1}{2 \ell} (\gamma_{a} Q)_\alpha & \lbrace Q_{\alpha}, Q_{\beta} \rbrace &= \big( \gamma^{a} \big)_{\alpha \beta} P_{a} + \frac{1}{\ell} \big( \gamma_\ast \big)_{\alpha \beta} K
\end{align}
\end{subequations}
where $P_0$, $P_1$ are the generators of time- and space-translations, $K$ generates Lorentz boosts, and $Q_{\alpha}$ are the Grassmann-valued fermionic supercharges. We define the gamma matrix appearing in the chiral projectors as $\gamma_\ast := \gamma^0\,\gamma_1$, see Appendix \ref{Appendix:Notation} for our conventions.

We are interested in the Carroll version of this algebra. We implement the corresponding {\.I}n\"on\"u--Wigner contraction by rescaling the generators of time translations and boosts linearly in the speed of light $c$ without rescaling the generator of spatial translations. 
\begin{equation}\label{rescal}
H= c \,  P_0   \qquad\qquad  B = c \, K  \qquad\qquad P = P_1\,.
\end{equation}
Here $H$ and $P$ are the new generators of time and space translations respectively, and $B$ generates the Carroll boosts. To retain the anticommutator of the supercharges proportional to the Hamiltonian, we rescale them as
\begin{equation}
\widehat{Q}_{\alpha} = \sqrt{c}\, Q_{\alpha}\,.
\end{equation}
Taking the limit of vanishing speed of light, $c\to 0$, and dropping the hat on $Q_\alpha$, we get the non-vanishing (graded) commutators generating the $\mathcal{N}=1$ Carroll--AdS$_2$ superalgebra
\begin{subequations}
\label{calgebra}
\begin{align}
\big[ B, P\big] &= H & \big[ H, P \big] &= \frac{1}{\ell^2}\, B  \\
\big[ P, Q_\alpha \big] &= \frac{1}{2\ell} \, (\gamma_1 Q)_\alpha
& \lbrace Q_{\alpha}, Q_{\beta} \rbrace &= \big( \gamma^{0} \big)_{\alpha \beta}\, H + \frac{1}{\ell} \,\big( \gamma_\ast \big)_{\alpha \beta} \, B\,.
\end{align}
\end{subequations}

Throughout this paper, we only use upper indices for $\gamma^0$ and lower ones for $\gamma_1$. The reasoning is as follows: although we start from a Lorentzian representation of the $\gamma$ matrices when performing the contraction, we expect the Carrollian results to be expressible in terms of Carrollian $\gamma$ matrices. While the Carrollian versions of $\gamma_0$ and $\gamma^1$ are degenerate and differ from their Lorentzian counterparts, the gamma matrices $\gamma^0$, $\gamma_1$, and $\gamma_\ast$ are the same in the Lorentzian and Carrollian cases. For more details see Appendices \ref{app:2} and \ref{app:3}.

\subsubsection{BF formulation}

We formulate sCJT as a BF theory 
\begin{equation}
S_{BF} [\mathcal{X}^\ast,A] = \frac{\kappa}{2 \pi} \int_{\mathcal{M}} \mathcal{X}^\ast \, F
\end{equation}
based on the AdS$_2$ superalgebra \eqref{algebra}. Here $\kappa$ is a dimensionless constant, $\mathcal{X}^\ast = X^I E_I$ is a scalar transforming in the co-adjoint representation of the Lie superalgebra $g$ on which the theory is based, and the Lie superalgebra valued 1-form $A = (A_\mu \extd x^\mu)_I \, E^I $ is a gauge field with the non-abelian field strength $F = \extd A + \frac12\,[A \stackrel{\wedge}{,}\, A]$. The structure constants $C_{K}{}^{I J}$ of $g$ are defined by
\begin{equation}
\big[E^I, E^J \big] = C_{K}{}^{I J} \, E^{K}\,.
\end{equation}
The dual $g\ast$ has a basis $E_I$ given by
$E_I (E^J) = \delta_J^I$. Then, the BF Lagrangian yields
\eq{
\mathcal{L}=\mathcal{X}^\ast \, F = X^K \big( \extd A_K + \frac{1}{2} C_{K}{}^{I J} A_I \wedge A_J \big)\,.
}{eq:lalapetz}

To build the sCJT action, we start by expanding the gauge field 
\begin{equation}\label{1form}
 A  = \tau \, H + e \, P + \omega \, B + \bar{Q} \, \Psi
\end{equation}
or, equivalently, in index notation, $A_I = (\tau,e,\omega, \Psi_{\alpha})$ and $E^I=(H,P,B,Q^{\alpha})$. In the context of the Einstein--Cartan formalism we can interpret $\tau$ and $e$ as the clock 1-form and spatial einbein 1-form, respectively, $\omega$ as the connection associated with local Carroll boosts, and $\Psi_{\alpha}$ as the Rarita--Schwinger 1-form. The gauge field $A$ transforms under gauge transformations in the usual non-abelian way,
\begin{equation}\label{BFtf}
\delta_{\epsilon} A_{I} = 
\extd \epsilon_{I} - C_I{}^{K J} \epsilon_K A_J 
\end{equation}
where $\epsilon_I$ is the parameter of the transformation. For the Carroll AdS$_2$ algebra \eqref{calgebra}, the variations of the gauge field along $\epsilon_I=(\lambda_H,\lambda_P,\lambda,\epsilon_{\alpha})$ yields
\begin {subequations}
\label{tfs}
\begin{align}
\delta_{\epsilon} \tau &= \extd \lambda_H + \lambda_P \, \omega - \lambda \, e - \bar{\epsilon} \gamma^0 \Psi\\
\delta_{\epsilon} e &= \extd \lambda_P\\
\delta_{\epsilon} \omega &= \extd \lambda + \frac{1}{\ell^2} \lambda_P \tau - \frac{1}{\ell^2} \lambda_H \, e - \frac{1}{\ell} \bar{\epsilon} \gamma_\ast \Psi\\
\delta_{\epsilon} \Psi &= \extd \epsilon -  \frac{1}{2 \ell}  \gamma_1 \epsilon \, e + \frac{1}{2 \ell} \lambda_P  \gamma_1 \Psi\,.
\end{align}
\end{subequations}

The curvature two-form splits as
\begin{subequations}
\begin{align}
F_H &= \extd \tau + \omega \wedge e + \frac{1}{2} \bar{\Psi} \gamma^0 \wedge \Psi\\
F_P &= \extd e\\
F_B &= \extd \omega + \frac{1}{\ell^2} \tau \wedge e + \frac{1}{2 \ell} \bar{\Psi} \gamma_\ast \wedge \Psi\\
F_Q &= \extd \Psi - \frac{1}{2 \ell} e \wedge \gamma_ 1 \Psi\,.
\end{align}
\end{subequations}
The scalar $\mathcal{X}^\ast$ can be expanded in the dual basis $g\ast$ as
\begin{equation}\label{scalar}
 \mathcal{X}^\ast  = \XH \, H^\ast + \XP \, P^\ast + X \, B^\ast + \bar{\chi} \, Q^\ast
\end{equation}
with $X^{I} = (\XH,\XP,X,\chi^{\alpha})$ and $E_I=(H^\ast,P^\ast,B^\ast,Q^\ast_{\alpha})$. As the Carroll AdS$_2$ superalgebra \eqref{calgebra} does not admit a non-degenerate symmetric invariant bilinear form $\langle \cdot , \cdot \rangle: g \times g \rightarrow \mathbb{R}$, which could be used to define the inverse $X_I = g_{IJ} X^J$, we do not have a metric BF theory. However, this is not a mandatory requirement for defining a BF theory, since the Lagrangian \eqref{eq:lalapetz} contracts coadjoint scalars with the adjoint field strength,
\begin{equation}
 \mathcal{L} = X \, F_B + \XH \, F_H + \XP \, F_P + \bar{\chi} \, F_Q\,.
\end{equation}
Inserting the expressions for curvature explicitly, the sCJT action expands as
\begin{align}\label{JT}
\nonumber
S_{\textrm{\tiny sCJT}} = \frac{\kappa}{2 \pi} \int_{\mathcal{M}} &\Big(X \big( \extd \omega + \frac{1}{\ell^2} \tau \wedge e + \frac{1}{2\ell} \bar{\Psi} \gamma_\ast \wedge \Psi \big) + \XH \big( \extd \tau + \omega \wedge e + \frac{1}{2} \bar{\Psi} \gamma^0 \wedge \Psi \big) + \XP \extd e \\
&+ \bar{\chi} \, \big( \extd \Psi - \frac{1}{2 \ell} e \wedge \gamma_ 1 \Psi \big)\Big)
\end{align}
where we identify $X$ as the dilaton, $\XH$ and $\XP$ as Lagrange multipliers for torsion and intrinsic torsion constraints, and $\chi$ as the dilatino. The action \eqref{JT} is invariant under the supersymmetry transformations \eqref{tfs} along the fermionic parameter $\epsilon$, together with
\eq{  
\delta_{\epsilon} X = \delta_{\epsilon} X_{H} = 0\qquad\qquad
\delta_{\epsilon} \XP =  \frac{1}{2 \ell} \bar{\chi} \gamma_1 \epsilon \qquad\qquad
\delta_{\epsilon} \bar{\chi} = \frac{1}{\ell} X \bar{\epsilon}\gamma_\ast + \XH \bar{\epsilon}\gamma^0\,.
}{tfs2}

The sCJT action \eqref{JT} can also be obtained from a Carrollian limit of the relativistic supersymmetric JT model given in \cite{Cardenas:2018krd}, along the lines of \cite{Grumiller:2020elf,Gomis:2020wxp}. In our conventions, the Lorentzian supersymmetric JT action 
\begin{equation}
S_{\textrm{\tiny sJT}} = \frac{\kappa}{2\pi} \int  \big(X^0 F_0 + X^1 F_1 + X F_{\Omega} + \bar{\chi} F_{\Phi} \big)
\end{equation}
entails the field strengths
\begin{subequations}
\begin{align}
F_0 &= \extd e_0 + \Omega \wedge e_1 + \frac{1}{2} \bar{\Phi} \wedge \gamma^0  \Phi\\
F_1 &= \extd e_1 + \Omega \wedge e_0 + \frac{1}{2} \bar{\Phi}  \wedge \gamma_1 \Phi\\
F_{\Omega} &= \extd \Omega + \frac{1}{{\ell}^2} e_0 \wedge e_1 + \frac{1}{2 \ell} \bar{\Phi} \wedge \gamma_\ast \Phi\\
F_{\Phi} &= \extd \Phi - \frac{1}{2} \Omega \wedge \gamma_\ast \Phi + \frac{1}{2 \ell} e_0 \wedge \gamma^0 \Phi - \frac{1}{2 \ell} e_1 \wedge \gamma_1 \Phi \,.
\end{align}
\end{subequations}
Rescaling the 1-forms as
\begin{equation} \label{elimit1}
\Omega = c \, \omega \qquad\qquad e_0= c \,  \tau \qquad\qquad e_1 = e \qquad\qquad \Phi = \sqrt{c}\, \Psi
\end{equation}
and considering an analogous expansion in $c$ on the target space coordinates
\begin{equation}\label{elimit2}
X \rightarrow c \, X  \qquad\qquad X^0 \rightarrow c \, \XH  \qquad \qquad X^1 \rightarrow c^2 \, \XP \qquad\qquad \chi \rightarrow c^{3/2} \chi 
\end{equation}
we truncate the action at order $\mathcal{O}(c^2)$. This turns out to be, in Carroll jargon, the so-called magnetic limit of Lorentzian 2d dilaton supergravity, which is obtained after rescaling $\kappa \rightarrow \kappa/c^2$ and then taking the limit $c\rightarrow 0$. The resulting action coincides precisely with \eqref{JT}. For a more detailed discussion on how to perform the magnetic Carroll limit, and its distinction to the electric one, see, e.g., \cite{Hansen:2021fxi,Campoleoni:2022ebj,Ecker:2023uwm,Bergshoeff:2023vfd}.

\subsubsection{gPSM formulation}

To deform the sCJT action \eqref{JT} to the most general 2d Carroll dilaton supergravity model, we express it as a gPSM (see, e.g., \cite{Ertl:2000si}) and then use the rigidity of gPSMs \cite{Izawa:1999ib}, which are 2d topological, non-linear gauge theories \cite{Ikeda:1993fh,Schaller:1994es}.

For the construction of the gPSM we take a 2d base supermanifold $\mathcal{M}_2$ with coordinates $x^{\mu}$, and define a target superspace $\mathcal{T}$ whose coordinates $X^I(x)=(X,\XH,\XP,\chi^{\alpha})$ are functions of the base manifold coordinates. As we begin dealing with the $\mathcal{N}=1$ case, here we only consider one pair of fermionic coordinates, but its extension to higher dimensions is always possible. We associate the scalar functions $X^I(x)$ with certain gauge fields $A_I(x) =(\omega, \tau, e, \Psi_{\alpha})$ whose components can be viewed as 1-forms on the base manifold $\mathcal{M}_2$ with values in the cotangent space of $\mathcal{T}$. The gPSM bulk action
\begin{equation}\label{PSM}
S_{\textrm{\tiny gPSM}} [X^I,A_I] = \frac{\kappa}{2 \pi} \int_{\mathcal{M}_2} \big(X^I \, \extd A_{I} + \frac{1}{2} P^{IJ}(X^K) A_I \wedge A_J \big)
\end{equation}
entails the Poisson tensor $P^{IJ}(X^K)$, which encodes the desired symmetries and the dynamics of the theory. In the graded case it satisfies graded anti-symmetry
\begin{equation}\label{ant}
P^{IJ} = - (-1)^{IJ} P^{JI}
\end{equation}
and graded non-linear Jacobi identities
\begin{equation}\label{jacobi}
P^{LI} \partial_L P^{JK} + (-1)^{J(I+K)} P^{LJ} \partial_L P^{KI} + (-1)^{I(K+J)} P^{LK} \partial_L P^{IJ} = 0\,.
\end{equation}
We use the convention that in the expression $(-1)^I$ the indices are $I=0$ for bosons and $I=1$ for fermions. The condition \eqref{ant} tells us that the Poisson tensor is anti-symmetric in the boson and mixed boson-fermion indices, and symmetric in the pure fermionic ones. Since the bosonic sector of the Poisson tensor is anti-symmetric and hence must have an even rank, it necessarily has a non-trivial kernel since its bosonic dimension is odd. Physically, this kernel corresponds to a conserved Casimir that can be interpreted as mass, as we shall see in Section \ref{solutions}.  

The variations of $X^I$ and $A_I$ in \eqref{PSM} yield the EOM
\begin{subequations}
\label{eofpsm}
\begin{align}
\delta A_I &: & \extd X^I - P^{JI} A_J &= 0\\
\delta X^I &: & \extd A_I + \frac{1}{2} \partial_I P^{JK} \, A_J \wedge A_K &= 0\,.
\end{align}
\end{subequations}
These are first-order PDEs for the fields $X^I(x)$ and $A_I(x)$. The gPSM action \eqref{PSM} is invariant under the non-linear gauge symmetries
\begin{equation}\label{PSMtf}
\delta X^I = P^{JI} \epsilon_J \qquad \qquad \delta A_I = \extd \epsilon_I - (\partial_I P^{JK}) \epsilon_J \, A_K \,.
\end{equation} 
When the Poisson tensor is linear in the target space coordinates, its partial derivatives $\partial_I P^{JK}$ result in the structure constants of the Lie superalgebra $C_I{}^{JK}$, so the right side of equation \eqref{PSMtf} turns into \eqref{BFtf}. Hence, the gPSM is a generalization of non-abelian BF theories to non-linear gauge theories. 

Furthermore, different gPSM models can be related by a diffeomorphism on the target space coordinates of the type $X^I \rightarrow \hat{X}^I (X^J)$, in which case the Poisson tensor transforms according to
\begin{equation}\label{poisdiff}
P^{IJ} (X^A) = (-1)^{KI+JI+KL+K} \, \,  \frac{\partial X^I}{\partial \hat{X}^L}  \, \hat{P}^{LK} (\hat{X}^A) \frac{\partial X^J}{\partial \hat{X}^K}\,.
\end{equation}
This transformation law is a general feature of PSM models and very useful, as we will show later, for interpolating between solution spaces of diffeomorphic models.

Each Poisson tensor defines a different theory; in our case of interest, the sCJT action \eqref{JT} can be expressed as a gPSM with the Poisson tensor
\begin{align}\label{poisJT}
&P^{X \XP} = \XH \qquad \qquad P^{\XH \XP} = \frac{1}{\ell^2} X \qquad \qquad P^{\XP \alpha} = -\frac{1}{2 \ell} \chi^{\beta}\big(\gamma_1 \big)_{\beta}^{\, \, \alpha} \\
\nonumber
&P^{\alpha \beta} = - \frac{X}{\ell} \big(\gamma_\ast \big)^{\alpha \beta} - \XH \big(\gamma^0 \big)^{\alpha \beta}\,.
\end{align}
The fermionic transformations \eqref{tfs} and \eqref{tfs2} can alternatively be obtained from \eqref{PSMtf}. From this formulation it can be seen that the $P^{X I}$ components of the Poisson tensor assure the local Carroll boost invariance of the theory when a boost transformation $\delta_{\lambda} \omega = \extd \lambda$, $\delta_{\lambda} \tau = - e \lambda$, $\delta_{\lambda} \XP = \XH \lambda$ is implemented.

\subsection{gPSM deformation}\label{se:2.2}
Consistent deformations allow to deform the gauge symmetries, without altering the number of field- or gauge degrees of freedom. Thus, consistent deformations maintain the number of local physical degrees of freedom (see, e.g., \cite{Berends:1984rq,Barnich:1993vg,Gomis:1995jp,Barnich:2000zw}). In \cite{Izawa:1999ib}, the rigidity of PSM models was proven, which implies that the most general consistent deformation of a linear BF theory is a nonlinear PSM model with the same dimension of the target space. An analogous statement applies to gPSMs. Thus, the most general consistent deformation of sCJT gravity is a generic gPSM with the same dimension of the target superspace as the sCJT model. Imposing that the deformed theory is still SUSY Carroll gravity imposes further constraints that we explore in this Subsection.

We are going to construct the most general consistent deformation of sCJT gravity by generalizing some components of the Poisson tensor to non-linear functions of the target space coordinates, while demanding the model to preserve local Carroll and SUSY symmetries. To achieve the former, as pointed out previously, we must keep fixed the $P^{X I}$ to assure local Carroll boost invariance. On the other hand, the criteria for assuring the supersymmetry of the theory (i.e., on how to fix the fermionic components of $P^{IJ}$) will be left for discussion in the following Sections. The most general deformation of \eqref{poisJT} to a generic gPSM describing Carrollian dilaton supergravity is then given by a Poisson tensor of the form
\begin{equation}\label{defpsm}
P^{IJ}=
\begin{bmatrix}
0 & 0 & \XH & 0 \\
0 & 0 & \mathcal{V} & P^{\XH \alpha} \\
-\XH & -\mathcal{V} & 0 & P^{\XP \alpha} \\
0 & - P^{\XH \alpha} & - P^{\XP \alpha} & P^{\alpha \beta} \\
\end{bmatrix}
\end{equation}
where we identify $\mathcal{V}$ as the bosonic generating potential of the theory, $P^{\XH \alpha}$ and $P^{\XP \alpha}$ as the mixed terms that relate the coupling between the fermionic sector and the temporal/ spatial einbein and $P^{\alpha \beta}$ as the purely fermionic $2 \times 2$ symmetric component of the Poisson tensor. At this stage, all of them are arbitrary functions of the target space coordinates $(X,\XH,\XP,\chi^{\alpha})$; however, they still need to be compatible with the graded Jacobi identities \eqref{jacobi}.

\subsubsection{Constraints from Jacobi identities}
We impose the graded Jacobi identities \eqref{jacobi} on the new Poisson tensor \eqref{defpsm}. As we have three antisymmetric and two symmetric components, this yields $15$ constraints on the tensor. The non-trivial constraints restrict the dependence of the free functions
\begin{align}
\mathcal{V} &= \mathcal{V}(X,\XH,\chi)  &P^{\alpha \beta} &= P^{\alpha \beta} (X,\XH,\chi)\\
P^{\XH \alpha} &= g^{\alpha}(X,\XH,\chi) &P^{\XP \alpha} &= f^{\alpha}(X,\XH,\chi)
\end{align}
together with
\begin{align}\label{constraint00}
P^{(\alpha \sigma} \partial_{\sigma} g^{\beta)} &= 0 \\
\label{constraint0}
P^{\alpha \beta} \partial_{\beta} \mathcal{V} - \XH \, \partial_X g^{\alpha} - \partial_{\XH} (\mathcal{V} \, g^{\alpha}) + f^{\beta} \partial_{\beta} g^{\alpha} &= 0 \\
\label{constraint}
\mathcal{V} \, \partial_{\XH} P^{\alpha \beta} + \XH \partial_X P^{\alpha \beta} + 2 P^{(\alpha \sigma} \partial_{\sigma} f^{\beta)} - f^{\sigma} \partial_{\sigma} P^{\alpha \beta} + 2 g^{(\alpha} \partial_{\XH} f^{\beta)}&=0\,.
\end{align}

We see that the constraints are compatible with $P^{\alpha \beta}$ being a symmetric tensor. Given a bosonic potential $\mathcal{V}$, the full Poisson tensor is not entirely determined by these equations, which is expected from comparison with the Lorentizan case. Therefore, we need to consider further restrictions for the deformed supergravity, following the prescriptions of \cite{Ertl:2000si} for a genuine supersymmetric extension. Before that, it is convenient to work on the functional form of the Poisson tensor explicitly to decouple the soul ($\sim \chi^2$ dependence) and body (pure bosonic) contributions. We start by expanding its components according to invariant quantities, $\gamma$ matrices and taking into account that due to the anticommutativity of $\chi^{\alpha}$, the dependence on $\chi^2:=\bar{\chi}\chi$ is at most linear. We denote the body components with hats and the soul ones with bars. This yields
\begin{subequations}
\begin{align}
V &= \hat{\mathcal{V}}(X, \XH) + \chi^2 \, \bar{\mathcal{V}}(X,\XH)\\
P^{\alpha \beta} &= U_0(X,\XH,\chi^2) \, \big(\gamma^0 \big)^{\alpha \beta} + U_1 (X,\XH,\chi^2) \, \big(\gamma_1 \big)^{\alpha \beta} + U_\ast (X,\XH,\chi^2) \, \big(\gamma_\ast \big)^{\alpha \beta}
\end{align}
\end{subequations}
where it is also convenient to use a combined notation for the $\chi^2$ dependence of $P^{\alpha \beta}$,
\begin{equation}
P^{\alpha \beta} = \hat{P}^{\alpha \beta} \, [\hat{U}_0,\hat{U}_1,\hat{U}_\ast] + \chi^2 \, \bar{P}^{\alpha \beta} \, [\bar{U}_0,\bar{U}_1,\bar{U}_\ast]\,.
\end{equation}

Notice that we do not include the identity matrix in the expansion of $P^{\alpha \beta}$; there is no loss of generality in that because, as we deal with a symmetric tensor of two dimensions, there are only three generators of the algebra, and we take the $\gamma$-matrices as a basis. For the mixed components of the tensor, the fermionic index structure and the anticommutativity of the dilatino fields only allows for the expansion 
\begin{equation}\label{fexpans}
f^{\alpha} = \chi^{\beta} F_{\beta}^{\, \, \alpha}(X,\XH), \qquad \qquad  g^{\alpha} = \chi^{\beta} G_{\beta}^{\, \, \alpha}(X,\XH)\,.
\end{equation}

The constraint \eqref{constraint00} leaves us with only two possibilities: either we take $\partial_{\beta}g^{\alpha} = 0$ or we demand each term of $P^{\alpha \beta}$ to vanish. But, as we will discuss in the next Section, we need to keep fixed $\hat{U}_0 $ to a certain non-zero value to preserve supersymmetry, so at least one component will be necessarily non-vanishing. With that consideration, and taking into account the fermionic expansion \eqref{fexpans}, the only suitable choice is to take $G_{\beta}^{\, \, \alpha} = 0$. Hence,
\begin{equation}
P^{\XH \alpha} = 0
\end{equation} 
preserves the vanishing commutator $[H,Q]=0$ of the Carroll algebra after any deformation. Because of this, the constraint \eqref{constraint0} gets simplified considerably and, following an analogous argument for the non-vanishing components of $\hat{P}^{\alpha \beta}$, lets us take $\bar{\mathcal{V}}=0$. Thus, in contrast to the Lorentzian case, the Carroll theory loses any possible soul contribution to the bosonic potential.

To solve the last constraint \eqref{constraint}, we expand the spinor matrix $F_{\beta}^{\, \, \alpha}$ in terms of the linearly independent $\gamma$ matrices and obtain the identity
\begin{equation}
F_{\beta}^{\, \, \alpha} = f(X,\XH) \delta_{\beta}^{\, \, \alpha} + f_0(X,\XH) \, \big(\gamma^0 \big)_{\beta}^{\, \, \alpha} + f_1(X,\XH) \, \big(\gamma_1 \big)_{\beta}^{\, \, \alpha} + f_\ast(X,\XH) \, \big(\gamma_\ast \big)_{\beta}^{\, \, \alpha}\,.
\end{equation}
Then, we solve the body and soul sectors of \eqref{constraint} separately, which yields
\begin{align}\label{bar}
&2 f \, \bar{P}^{\alpha \beta} - \bar{P}^{\alpha \sigma} F_{\sigma}^{\, \, \beta} - \bar{P}^{\beta \sigma} F_{\sigma}^{\, \, \alpha} - \XH \, \partial_X \bar{P}^{\alpha \beta} - V \, \partial_{\XH} \bar{P}^{\alpha \beta} = 0\\ \label{hat}
& \hat{P}^{\alpha \sigma} F_{\sigma}^{\, \, \beta} + \hat{P}^{\beta \sigma} F_{\sigma}^{\, \, \alpha} + \XH \, \partial_X \hat{P}^{\alpha \beta} + V \, \partial_{\XH} \hat{P}^{\alpha \beta} = 0 \,.
\end{align}
Using again the independence of the $\gamma$ matrices on \eqref{hat}, we get for the body sector
\begin{align}\label{hateq}\nonumber
&\big( \XH \, \partial_X + \mathcal{V} \, \partial_{\XH}) \frac{\hat{U}_0}{2} = \hat{U}_1 f_\ast - \hat{U}_\ast f_1 -  \hat{U}_0 f\\
&\big(\XH \, \partial_X + \mathcal{V} \, \partial_{\XH}) \frac{\hat{U}_1}{2} = \hat{U}_0 f_\ast - \hat{U}_\ast f_0 - \hat{U}_1 f\\
\nonumber
&\big( \XH \, \partial_X + \mathcal{V} \, \partial_{\XH}) \frac{\hat{U}_\ast}{2} = \hat{U}_1 f_0 - \hat{U}_0 f_1 - \hat{U}_\ast f
\end{align}
and an analogous expression for the soul, which shows that both sectors are decoupled from each other. At this stage, we have been working with complete generality and without further assumptions on the form of the fermionic tensor. In the next Subsections, we shall consider a simplified family of models where the bosonic potential does no depend on $\XH$ and then deform it to the most general case by applying a diffeomorphism on the target space coordinates.

\subsubsection{Dilaton prepotential family}
Let us first consider a family of simplified models where the bosonic potential is restricted to be a function of the dilaton field $X$ only, which allows us to write $\mathcal{V}(X,\XH) = V(X)$. As discussed in \cite{Ertl:2000si}, in this case, the limit of rigid supersymmetry in flat space, along with other requirements for a genuine supergravity deformation, constrains the value of the purely fermionic part of the Poisson tensor: it demands $U_0$ and $U_1$ to be fixed to coincide with the rigid limit (that is, when the deformation is turned off and curvature vanishes) and to consider $U_\ast$ only to depend on $X$. For later convenience, we define $U_\ast:= u(X)/2$, where the factor $2$ is conventional. This ansatz avoids singularities that could appear at certain loci of the target space when realizing the supersymmetric extension. Furthermore, here and in the following Sections we work within the full fermionic rank supergravity, i.e., we take the $P^{\alpha \beta}$ for rigid supersymmetry to be non-degenerate. 

From taking $\ell \rightarrow \infty$ in \eqref{poisJT}, we obtain the rigid supersymmetry limit, which let us fix $U_0 = - \XH$ and $U_1 = 0$. Then, by solving the soul equations with this ansatz, we get that all contribution must vanish $\bar{U}_0 = \bar{U}_1 = \bar{U}_\ast = 0$. So, the fermionic sector can be written as $U_i=\hat{U}_i$. To fix an implicit relation between the so far non-related potentials $\mathcal{V}$ and $U_\ast$, we still have enough freedom to demand $f_\ast$ to vanish. In \cite{Ertl:2000si} it was shown that this choice eliminates singular terms at $\XH=0$ and makes the Poisson tensor regular. Hence, using \eqref{hateq} we get that $f_0=0$, which is an implicit consequence of the Carrollian symmetry of the theory, along with 
\begin{align}\label{const}
\frac{u(X)}{2} \, f_1 - \XH \, f &= \frac{V(X)}{2}\\\label{const2}
V(X) \, \XH + \Big(2 \XH^2 - \frac{u^2}{2} \Big) \, f &=   \XH \, \partial_X  \frac{u^2}{8}\,.
\end{align}
To ensure the functions remain regular, Eq.~\eqref{const2} requires eliminating the dependence on $\XH^2$, which in turn demands $f = 0$. Solving it provides a constraint between $u$ and $V$,
\begin{equation}\label{convex}
u^2(X) = 8 \int V(X) \extd X\,.
\end{equation}
 From this expression, we can identify $u(X)$ with the fermionic \textit{prepotential} given in \cite{Bergamin:2002ju}, which serves as a generator of the supersymmetric extension of dilaton theories. The deformed Poisson tensor of the prepotential family reads
\begin{align}\label{prepo}
&P^{X \XP} = \XH \qquad \qquad P^{\XH \XP} = \frac{(u^2(X))'}{8} \qquad \qquad P^{\XP \alpha} = \frac{u'(X)}{4} \chi^{\beta}\big(\gamma_1 \big)_{\beta}^{\, \, \alpha} \\
\nonumber
&P^{\alpha \beta} =  - \XH \big(\gamma^0 \big)^{\alpha \beta} + \frac{u(X)}{2} \, \big(\gamma_\ast \big)^{\alpha \beta}\,.
\end{align}
This family of models includes, as expected, the sCJT theory for the choice $u= -\frac{2X}{\ell}$, but it generalizes to a broader class of models. However, notice that, as in the Lorentzian case, the relation \eqref{convex} between bosonic potential $V$ and prepotential $u$ imposes a convexity condition on the former since both sides must be non-negative. This shows the incompatibility of this supersymmetric extension with certain models (for example, the dS version of JT where $V(X)=- \frac{X}{\ell^2}$ and hence no real-valued prepotential can be defined). Furthermore, there is no unique choice for the prepotential because is defined up to a change of sign $u \rightarrow -u$; the branch we use in this paper is in agreement with the literature.

\subsubsection{General deformed supergravity}\label{generaldif}
Most 2d dilaton theories can be written as a subclass of models whose bosonic potential allows for a dilaton kinetic term in the second-order formulation. This subclass is given by $\mathcal{V}(X,\XH) = V(X) - \frac{\XH^2}{2} \, Z(X)$. To obtain this family of models from the dilaton prepotential family discussed above, we build them from a particular target space diffeomorphism $X^I \rightarrow \hat{X}^I (X^J)$ compatible with both fermionic and Carroll symmetry constraints, that was shown in \cite{Bergamin:2002ju} to be equivalent to a conformal transformation on the metric. Here we consider a dilaton-dependant Weyl rescaling on \eqref{prepo} parametrized by $Q(X)$
\begin{gather}\label{Weyl}
e \rightarrow \hat{e} = e^{Q(X)} \, e
\end{gather}
that leaves the dilaton invariant, $X \rightarrow \hat{X} = X$. These Weyl rescalings are compatible with
the absence of intrinsic torsion ($\extd e=0$ is an EOM of the theory). Consistency with Carroll boosts demand $\tau$ to scale in the same way, 
\begin{gather}
\tau \rightarrow \hat{\tau} = e^{Q(X)} \, \tau\,.
\end{gather}
Additionally, if we want to retain the dynamical terms $\XH \, \extd\tau$ and $\XP \, \extd e$ after the transformation, the scalars must transform as well,
\begin{align}
\XH \rightarrow \hat{X}_H = e^{-Q(X)} \, \XH \\
\XP \rightarrow \hat{X}_P = e^{-Q(X)} \, \XP\,.
\end{align}
Another feature of our theory that we want to retain is the rigid supersymmetry limit. In that way, demanding the term $ \XH \, \bar{\Psi} \wedge \gamma^0  \Psi$ to be fixed under the conformal transformation leads to 
\begin{gather}
     \psi_\alpha \rightarrow \hat{\psi}_\alpha= e^{ \frac{1}{2} Q (X)} \psi_\alpha \qquad 
     \qquad \chi^\alpha \rightarrow \hat{\chi}^\alpha=e^{ -\frac{1}{2} Q (X)} \chi^\alpha\,. 
\end{gather}
Furthermore, we want to preserve the prepotential $u$ as an arbitrary function that generates our theories, so we also have to redefine it under the Weyl transformation
\begin{gather}
     u \rightarrow \hat{u}=e^{-Q (X)} u\,.
\end{gather}
As a result, the bosonic potential changes according to 
\begin{equation}
V \rightarrow \frac{1}{8} e^{-2Q(X)} ( (u^2)' - 2 Q^{\prime} u^2) \,.
\end{equation}
Consistency of the Lagrangian requires absorbing the remnant terms in the transformation law for the spin connection
\begin{gather} 
\omega \rightarrow \hat{\omega}=\omega + Q^{\prime} (\XH \tau+ \XP e+\frac{1}{2} \chi^\beta \psi_\beta)\,.
\end{gather}

By defining $Z(X) = -2 Q'(X)$ after the transformation, we obtain the general deformed Poisson tensor,
\begin{align}\label{poisdef}
P^{X \XP} &= \XH  & P^{\XH \XP} &= V(X) - \frac{\XH^2}{2} \, Z(X) \\
\nonumber
P^{\XP \alpha} &= \frac{V}{u} \chi^{\beta}(\gamma_1)_{\beta}^{\, \, \alpha} + \frac{Z}{4} \XH \chi^{\beta} \, \delta_{\beta}^{\, \, \alpha} & P^{\alpha \beta} &=  - \XH (\gamma^0)^{\alpha \beta} + \frac{u(X)}{2} \, (\gamma_\ast)^{\alpha \beta}
\end{align}
with the bosonic potential
\begin{equation}\label{potential}
V(X) = \frac{1}{8} \big( (u^2)' + u^2 Z \big) \,.
\end{equation}
We can see that dilaton-dependent Weyl rescalings \eqref{Weyl} act in the same way as in
the Lorentzian case and can be used to introduce or eliminate a kinetic potential
function $Z(X)$ into the full bosonic potential $\mathcal{V}(X,\XH)$. Notice that the result \eqref{poisdef} could also be obtained 
by applying a diffeomorphism on the target space coordinates. With the coordinate change
\begin{equation}
X^I (\hat{X}^J) = \big(\hat{X}, e^{Q(\hat{X})} \hat{X}_H, e^{Q(\hat{X})} \hat{X}_P, e^{\frac{1}{2}Q(\hat{X})} \hat{\chi}^{\alpha} \big)
\end{equation}
the transformed components of the graded Poisson tensor can be computed with \eqref{poisdiff} and yield the same result \eqref{poisdef}.

\subsubsection{Limit from the Lorentzian theory}
Here, we show that the general deformed theory \eqref{poisdef} can also be obtained from taking the Carrollian limit of the Lorentzian version of the action. A subclass of 2d dilaton models with $\mathcal{V}(X,\XH) = V(X) - \frac{\XH^2}{2} \, Z(X)$ was built for local Lorentzian symmetries in \cite{Ertl:2000si}. Fitting their expression within our conventions, the action is
\begin{align}\nonumber
&S_{\textrm{\tiny sDIL}}^{L} = \frac{\kappa}{2 \pi} \int_{M} \Big[X \extd \omega + \Big(V(X) + \frac{X^a X_a}{2} \, Z  - \chi^2 \, \big(\frac{V \, Z + V}{2u} - \frac{2V^2}{u^3} \big)\Big) \frac{\epsilon_{ab}}{2} \, e^a \wedge e^b  \\
\nonumber
&+ \bar{\chi} \Big(\extd\Psi - \frac{1}{2} \omega \wedge \gamma_\ast \Psi + \frac{V}{u} \gamma^a e_a \wedge \Psi + \frac{Z}{4} \, X^a \gamma_a \gamma^b e_b \, \wedge \gamma_\ast \Psi \Big) -\frac{1}{4} \Big( u + \chi^2 \, \frac{Z}{4} \Big) \bar{\Psi} \, \gamma_\ast \wedge \Psi  \\
&  + X^a \Big( \extd e_a -\epsilon_{ab} \, \omega \wedge e^b + \frac{1}{2} \bar{\Psi} \, \gamma_a \wedge \Psi \Big)\Big]
\label{eq:angelinajolie}
\end{align}
with
\begin{equation}
V = \frac{1}{8} \big( (u^2)' + u^2 Z \big) \,.
\end{equation}
As we did in the JT case, we can take the leading order limit in the small $c$-expansion to obtain the Carrollian contraction of this theory. For that purpose, we consider the expansion we did in \eqref{elimit1}, \eqref{elimit2} 

\begin{align}\label{limitscaling}
\omega &\rightarrow c \, \omega & e^0 &\rightarrow c \, \tau & e^1 &\rightarrow e & \Psi &\rightarrow c^{1/2} \, \Psi \\
\nonumber
X &\rightarrow c \, X & X^0 &\rightarrow - c \, \XH & X^1 &\rightarrow c^2 \XP & \chi &\rightarrow c^{3/2} \chi 
\end{align}
and we demand the bosonic potential $\mathcal{V}(X,\XH)$ to rescale in a way that preserves the leading order terms in the expansion parameter, i.e., $\mathcal{V} \rightarrow c \, \mathcal{V}$. Thus, the prepotential functions must rescale as\footnote{A more rigorous derivation on how to apply this rescaling on arbitrary potential functions is performed in the $\mathcal{N} =2$ case, Section \ref{section}.} 
\begin{equation}
u \rightarrow c \, u \qquad Z \rightarrow \frac{1}{c} Z\,.
\end{equation}
Then, truncanting the action at lowest order in $c$ we get precisely \eqref{poisdef}, which matches with our Carrollian construction of the general deformed theory.

\subsection{Generalized Carroll dilaton supergravity}\label{se:2.3}
\subsubsection{First order formulation}
From the previous Section, we can write the generalized 2d Carroll dilaton supergravity first-order action as
\begin{align}\label{csDIL1st}
& S_{\textrm{\tiny sDIL}}^{C} = \frac{\kappa}{2 \pi} \int_{M} \Big[X  \extd \omega  + \XH \big( \extd \tau + \omega \wedge e + \frac{1}{2} \bar{\Psi} \wedge \gamma^0  \Psi \big) + \XP \, \extd e
\\\nonumber
& + \bar{\chi} \, \big( \extd \Psi + \frac{V}{u} e \wedge \gamma_ 1 \Psi + \frac{Z}{4} \XH \, e \wedge \Psi \big) + \big(V(X) - \frac{\XH^2}{2} \, Z(X) \big) \tau \wedge e - \frac{u(X)}{4} \bar{\Psi} \wedge \gamma_\ast  \Psi\Big]
\end{align}
where the dilaton dependant potential $V(X)$ and the prepotential $u(X)$ are related by \eqref{potential}. As before, the 0-forms are dilaton $X$, dilatino $\chi$ and
Lagrange multipliers for torsion constraints $\XH$, $\XP$. The 1-forms are spatial einbein $e$,
temporal einbein $\tau$, Carroll boost connection $\omega$ and gravitino $\Psi$. The composite 2-forms are curvature $\Omega = \extd \omega$, the fermionic partner of curvature $\sigma_{\alpha} = \extd \Psi_{\alpha}$, torsion $T = \extd\tau + \omega \wedge e$, and intrinsic torsion $\Theta = \extd e$; the latter is the part of the torsion that is independent of the boost connection. Due to the construction of the theory, the Lagrangian 2-form (and hence also the action) is invariant under local Carroll boosts,
\begin{align}\label{cboosts}
\delta_{\lambda} X &= 0 & \delta_{\lambda} \XH &=0 & \delta_{\lambda} \XP &= \XH \lambda & \delta_{\lambda}\chi &= 0\\
\nonumber
\delta_{\lambda} \omega &= \extd \lambda & \delta_{\lambda} \tau &= - e \, \lambda & \delta_{\lambda} e &= 0 & \delta_{\lambda} \Psi &= 0\,.
\end{align}
These transformations show that the dilaton $X$, the field $\XH$, and the spatial einbein
$e$ are Carroll boost invariant. The variations of the temporal einbein $\tau$ and the scalar $\XP$ procure that the torsion $T$ is a Carroll boost-invariant quantity. 

The action \eqref{csDIL1st} is also invariant under two additional bosonic gauge transformations with parameters $\lambda_H$ and $\lambda_P$ respectively,
\begin{align}
\delta_{\lambda_H} X  &=\delta_{\lambda_H} \XH=0  & \delta_{\lambda_H} \XP  &= \big(V - \frac{\XH^2}{2} Z \big) \lambda_H & \delta_{\lambda_H} \chi &= 0 
\label{lambdaH} \\\nonumber
\delta_{\lambda_H} \omega &= \big(\frac{\XH^2}{2} Z' -V' \big) e \lambda_H 
& \delta_{\lambda_H} \tau  &= \extd \lambda_H + \XH Z e \lambda & \delta_{\lambda_H} e &=  \delta_{\lambda_H} \Psi = 0\,.
\end{align}
and
\begin{align*}
\delta_{\lambda_P} X  &= -\XH \lambda_P  & \delta_{\lambda_P} \XH &= \big( \frac{\XH^2}{2} \, Z -V \big) \lambda_P & \delta_{\lambda_P} \XP &= 0 & \delta_{\lambda_P}\chi &= \Big(\frac{Z}{4} \XH \chi-\frac{V}{u} \gamma_1 \chi \Big) \lambda_P
\end{align*}
\begin{equation}
\delta_{\lambda_P} \omega = \Big(\big(V' - \frac{\XH^2}{2} \, Z' \big) \tau -\Big(\frac{V}{u} \Big)' \, \bar{\chi} \gamma_1 \Psi - \frac{Z'}{4} \, \XH \, \bar{\chi} \Psi\Big) \lambda_P
\label{lambdaP}
\end{equation}
\begin{align*}
\delta_{\lambda_P} \tau  &= \big( \omega - \XH Z \, \tau - \frac{Z}{4} \bar{\chi} \Psi \big)\lambda_P & \delta_{\lambda_P} e &= \extd \lambda_P & \delta_{\lambda_P} \Psi  &= \big( - \frac{V}{u} \gamma_1 - \frac{Z}{4}\big)\Psi \lambda_P \,.
\end{align*}

The local supersymmetric transformations are implemented by the fermionic parameter $\epsilon_{\alpha}$,
\begin{samepage}
\begin{align*}
\delta_{\epsilon} X  &= \delta_{\epsilon} \XH   = 0 &\delta_{\epsilon} \XP  &= -\frac{V}{u} \, \bar{\chi} \gamma_1 \epsilon - \frac{Z}{4} \XH \, \bar{\chi} \epsilon & \delta_{\epsilon} \chi  &= u \, \gamma_\ast \epsilon - \XH \, \gamma^0 \epsilon
\end{align*}
\begin{equation}\label{epsilon}
\mathclap{\delta_{\epsilon} \omega = u' \, \bar{\epsilon} \gamma_\ast \Psi + \Big(\frac{V}{u} \Big)^\prime \bar{\chi} \gamma_1 \epsilon \, e + \frac{Z'}{4} \, \XH \, \bar{\chi} \epsilon \, e}
\end{equation}
\begin{align*}
\delta_{\epsilon} \tau &= \bar{\epsilon} \gamma_0 \Psi + \frac{Z}{4} \bar{\chi} \epsilon \, e & \delta_{\epsilon} e &= 0 &
\delta_{\epsilon} \Psi  &= \extd \epsilon + \frac{V}{u} \, \gamma_1 \epsilon \, e + \frac{Z}{4} \XH \, \epsilon \, e\,.
\end{align*}
\end{samepage}

\subsubsection{Equations of motion}\label{eoms}
Varying the action \eqref{eq:angelinajolie} yields the EOM
\begin{align}\label{1}
	\delta \omega:&  \extd X+\XH \, e=0\\\label{2}
	\delta \tau:& \extd \XH +(V - \frac{\XH^2}{2} \, Z) \, e=0 \\\label{3}
	\delta e :&  \extd \XP -(V - \frac{\XH^2}{2} \, Z) \, \tau- \XH \, \omega+ \frac{V}{u} \bar{\chi} \gamma_1 \Psi+\frac{Z}{4}  \XH \, \bar{\chi} \Psi =0\\\label{4}
	\delta \Psi_\alpha :& \extd \chi^{\alpha} - \Big(\frac{u}{2} (\gamma_\ast)^{\alpha \beta} - \XH  (\gamma^0)^{\alpha \beta} \Big) \Psi_\beta -  \chi^\beta \Big( \frac{V}{u} (\gamma_1)_{\beta}^{\, \, \alpha} + \frac{Z}{4} \XH  \delta_{\beta}^{\, \, \alpha} \Big) \, e =0 \quad\\ \label{5}
	\delta X:&  \extd \omega + (V^{\prime}- \frac{\XH^2}{2} Z^{\prime}) \tau \wedge e - \frac{u^{\prime}}{4} \bar{\Psi} \wedge \gamma_\ast \Psi+ \Big( \frac{V}{u} \Big)^{\prime} \bar{\chi} e \wedge \gamma_1  \Psi+\frac{Z^{\prime}}{4} \XH \bar{\chi}   e \wedge \Psi=0 \\ \label{6}
	\delta \XH:& \extd \tau-\XH \, Z \, \tau \wedge e+ \omega \wedge e + \frac{1}{2} \bar{\Psi} \, \wedge \gamma^0 \Psi + \frac{Z}{4} \bar{\chi}  \, e \wedge \Psi=0 \\ \label{7}
	\delta \XP:& \extd e=0\\ \label{8}
	\delta \bar{\chi}^\alpha:&  \extd \Psi_\alpha +\frac{V}{u} \, e \wedge ( \gamma_1 )_{\alpha}^{\, \, \beta} \Psi_\beta + \frac{Z}{4}  \XH \, e \wedge \Psi_\alpha=0\,.
\end{align}
The first equation \eqref{1} allows algebraically determining the spatial einbein (and hence the Carroll metric) in terms of the Carroll boost invariant scalars, $X$ and $\XH$. The second equation \eqref{2} entails a conserved Casimir function, which we shall explicitly compute when discussing linear dilaton vacua solutions. The third equation \eqref{3} allows determining the auxiliary field $\XP$ in terms of the potentials $V(X)$ and $Z(X)$ and the geometric data extracted from the other EOM or, alternatively,
if $\XP$ is gauge fixed suitably it provides an algebraic constraint relating $\tau$, $\omega$, $\chi$ and $\Psi$. The fourth equation \eqref{4} can be used to determine the dilatino field $\chi_{\alpha}$ in terms of other quantities. The fifth equation \eqref{5} determines the Carroll curvature $\Omega = \extd \omega$, which generally is non-zero but trivially vanishes whenever the prepotentials $u$ and $Z$ are independent of the dilaton field. The sixth equation \eqref{6} shows that on-shell Carroll torsion $T = \extd \tau + \omega \wedge e$ is non-vanishing even for $Z=0$ due to the fermionic terms contribution. The seventh equation \eqref{7} reveals that there is never intrinsic torsion $\Theta = \extd e$, regardless of supergravity contributions or how the bosonic potential is chosen. The last equation \eqref{8} determines the fermionic counterpart of curvature $\sigma_{\alpha} = \extd \Psi_{\alpha}$ in terms of the gravitino field, and it can be used as a differential equation to solve for the latter.

\subsubsection{Second order formulation}
In this Section, we derive the second-order action for the general dilaton supergravity theory in 2d. We start by eliminating the dependence of the fields $\omega$ and $\XH$ from the first order action by using the EOM to integrate them out. The first step is to define the dual vectors of the einbeins $v^{\mu}$ and $e^{\mu}$ by
\begin{equation}\label{dual}
v^{\mu} \tau_{\mu} = -1  \qquad \qquad e^{\mu} e_{\mu} = 1 \qquad \qquad e^{\mu} \tau_{\mu} = v^{\mu} e_{\mu} = 0
\end{equation}
and use them to translate the EOM into algebraic expressions. Starting with equation \eqref{1}, it is solved by
\begin{equation}
\XH =- e^{\mu} \partial_{\mu} X
\end{equation}
with the constraint $v^{\mu} \partial_{\mu} X =0$. Next, it is useful to split $\omega = \hat{\omega} + t + \rho \, e$, where $\hat{\omega}$ is the torsionless part of the spin connection that satisfies $\extd \tau + \hat{\omega} \wedge e = 0$, a torsion part $t$ and $\rho$ an arbitrary function. The latter embodies the usual ambiguity that the Carrollian
spin connection is not entirely determined by the EOM \cite{Hartong:2015xda}. Plugging this ansatz into \eqref{6} yields
\begin{equation}
\hat{\omega}_{\mu} = - e^{\nu} \partial_{\mu} \tau_{\nu} + e^{\nu} \partial_{\nu} \tau_{\mu} = -2 e^{\nu} \partial_{[\mu} \tau_{\nu]}
\end{equation}
and 
\begin{equation}
t = \XH Z \, \tau + \frac{Z}{4} \chi \Psi + \hat{t} \qquad \text{with} \qquad \hat{t}_{\mu} = -e^{\nu} \bar{\Psi}_{\mu} \gamma^0 \Psi_{\nu}\,.
\end{equation}

Inserting these expressions for $\XH$ and $\omega$ into the first order action \eqref{csDIL1st} yields
\begin{align}\label{csDIL2nd}
&S = \frac{\kappa}{2 \pi} \int_{M} \Big[X\extd \hat{\omega} - \rho \extd X \wedge e + \Big(V - \frac{(e^{\mu} \partial_{\mu} X)^2}{2} Z \Big) \tau \wedge e  - \frac{(e^{\mu} \partial_{\mu} X)}{2} \bar{\Psi} \wedge \gamma^0  \Psi + \XP \extd e\nonumber\\
& + \bar{\chi} \, \big( \extd \Psi + \frac{V}{u} e \wedge \gamma_ 1 \Psi - \frac{Z}{4} (e^{\mu} \partial_{\mu} X) \, e \wedge \Psi \big) - \frac{u(X)}{4} \bar{\Psi} \wedge \gamma_\ast  \Psi\Big]\,.
\end{align}
It is convenient to work out a common basis $(\tau,e)$ for all the terms, so the next step is to consider a change of basis of the form
\begin{equation}
\extd x^{\mu} = - v^{\mu} \, \tau + e^{\mu} \, e
\end{equation}
so any 2-form $a = a_{\mu \nu} \extd x^{\mu} \wedge \extd x^{\nu}$, with $a_{\mu \nu}$ being antisymmetric, after the change of basis turns out to be $a = 2 a_{\mu \nu} e^{\mu} v^{\nu} \tau \wedge e$. Relating the torsionless connection with the Riemann curvature tensor by $R^{\lambda}_{\, \, \sigma \mu \nu} = - v^{\lambda} e_{\sigma} (\extd \hat{\omega})_{\mu \nu}$ and defining the curvature scalar in 2d as $R= 2 e^{\mu} e^{\nu} R^{\lambda}_{\, \, \mu \lambda \nu}$, we find $2 \extd\hat{\omega} = R \, \tau \wedge e$. Writing the fermionic 1-forms as $\Psi = \Psi_{\mu} \extd x^{\mu}$, we also get expressions for the dynamical terms of the action in our desired basis
\begin{equation}
\extd e = 2 e^{\mu} v^{\nu} \, \partial_{[\mu} e_{\nu]} \, \tau \wedge e := K \, \tau \wedge e \qquad\qquad \extd \Psi = 2 e^{\mu} v^{\nu} \, \partial_{[\mu} \Psi_{\nu]} \, \tau \wedge e := \sigma \, \tau \wedge e
\end{equation}
for $\sigma$ being the fermionic partner of the curvature scalar and $K$ identified as the trace of the extrinsic curvature. Defining the volume form $\tau \wedge e = \tau_{\mu} e_{\nu} \extd x^{\mu} \wedge \extd x^{\nu} = \text{det}(\tau, e) \extd x^2$, we finally arrive at the second order action
\begin{equation}
S_{2^{\textrm{\tiny nd}} \textrm{\tiny order}} = \frac{\kappa}{4 \pi} \int_{M} \big( \mathcal{L}_{\textrm{\tiny dil}} + \mathcal{L}_{f} \big) \, \text{det}(\tau,e) \, \extd x^2
\end{equation}
where the first term is the pure bosonic usual second order dilaton action for potentials $V(X)$ and $Z(X)$ 
\begin{equation}
\mathcal{L}_{\textrm{\tiny dil}} = XR + 2 \rho v^{\mu} \partial_{\mu} X + 2 \XP \, K + 2 V(X) - (e^{\mu} \partial_{\mu} X)^2 \, Z(X)
\end{equation}
with $\rho$ playing the role of a Lagrange multiplier for the above-mentioned constraint $v^{\mu} \partial_{\mu} X =0$, and the second term is the coupling to the fermionic sector of dilaton supergravity
\begin{equation}
\mathcal{L}_f = 2 \bar{\chi} \big( \sigma + \frac{V}{u} v^{\mu} \gamma_1 \Psi_{\mu} - \frac{Z}{4} (e^{\mu} \partial_{\mu} X) v^{\nu} \Psi_{\nu} \big) - 2 (e^{\sigma} \partial_{\sigma} X) e^{\mu} v^{\nu} \bar{\Psi}_{\mu} \gamma^0 \Psi_{\nu} - u(X) e^{\mu} v^{\nu} \bar{\Psi}_{\mu} \gamma_\ast \Psi_{\nu}\,.
\end{equation}
As discussed in \cite{Ecker:2023uwm}, the bosonic dilaton Lagrangian can be shown to be Carroll invariant using \eqref{cboosts} together with the dual transformations $\delta_{\lambda} v^{\mu}= 0$ and $\delta_{\lambda} e^{\mu}= - \lambda v^{\mu}$ that preserve \eqref{dual}. Furthermore, the Lagrange multipliers must transform as
\begin{equation}
\delta_{\lambda} \, \rho = - Z \lambda e^{\mu} \partial_{\mu} X + \nabla_{\mu} (e^{\mu} \lambda) \qquad \qquad \delta_{\lambda} \XP = - \lambda e^{\mu} \partial_{\mu} X
\end{equation}
where $\nabla$ is the connection associated with $\hat{\omega}$. For the fermionic Lagrangian $\mathcal{L}_f$, it is enough to use the antisymmetry of $\bar{\Psi}_{\mu} \gamma \Psi_{\nu}$ to show that each term is by itself Carroll invariant.

\subsubsection{Solutions of the EOM}\label{solutions}
For solving the EOM we follow the work done in \cite{Grumiller:2021cwg} and first consider the \textit{constant dilaton vacua}, which are solutions where $\XH$ vanishes everywhere. In that case, Eq.~\eqref{1} demands $\extd X =0$ and leads to a constant dilaton field $X_c$. Equation \eqref{2} demands that this constant cannot be anything but has to solve $V(X_c) = 0$. In particular, this means constant dilaton vacua need infinite finetuning of the dilaton field and may not even exist for some models (an example is the sCCJ model where $V(X) = \Lambda \neq 0$). Furthermore, the Carroll curvature is given by $\Omega = - V'(X_c) \, \tau \wedge e$, which results in a constant valued number times the volume form. As these solutions are not very rich in structure, we consider the generic sector, the one where $\XH$ only vanishes at specific isolated points or not at all. 

Let us assume we are solving the EOM in a patch where $\XH \neq 0$. Hence, we can use Eq.~\eqref{1} to solve for $e$,
\begin{equation}\label{lindil}
e = -\frac{\extd X}{\XH}\,.
\end{equation}
These solutions are called the \textit{linear dilaton vacua}. 

Our procedure for solving the remaining equations is to consider the dilaton prepotential case ($Z(X) \rightarrow 0$), and then return to the general deformed case by a Weyl rescaling. Taking that into account and inserting the expression for $e$ into Eq.~\eqref{2}, we get
\begin{equation}
\frac{1}{2} \extd(\XH)^2 - V(X) \extd X = 0
\end{equation}
which allows expressing $\XH$ as function of the dilaton $X$ and of an integration constant
$M$. We refer to $M$ as the Carrollian mass and it can be interpreted as the Carrollian Casimir function $M(X,\XH)$ with $\extd M=0$, which spans the kernel of the degenerate Poisson tensor \eqref{poisdef}. For computing this quantity, we define the integrated potential
\begin{equation}
w(X) := \int^X V(x) \, \extd x \,.
\end{equation}
So the conserved Carrollian mass yields
\begin{equation}\label{Cmass}
M(X,\XH) = w(X) - \frac{1}{2} \XH^2\,.
\end{equation}
In contrast to the supersymmetric Lorentizan theory (see for example \cite{Bergamin:2003am} and references therein), the Carrollian case has a vanishing soul contribution $\sim \chi^2$ of the Casimir function. This is in perfect agreement with the fact that both bosonic and fermionic potentials lose their souls (i.e., do not depend on the dilatino field) after the Carrollian limit performed by \eqref{limitscaling}. 

We now return to equation \eqref{7} which allows us to solve trivially for the spatial einbein by defining a Carroll radial coordinate $r$,
\begin{equation}
\extd e= 0 \qquad \Rightarrow \qquad e = \extd r\,.
\end{equation}
Inserting the Carrollian mass \eqref{Cmass} into \eqref{lindil} yields a differential equation for the dilaton,
\begin{equation}
\extd r = \mp \frac{\extd X}{\sqrt{2w(X) -2M}}
\end{equation}
where the two signs refer to the branches of the square root. The solution of this equation, if its inverse is well defined, allows to express the dilaton field in terms of the radial coordinate $X(r)$ and hence they can be used interchangeably.

To solve the graviton field we define some coordinate $t$ and use $(t,r)$ as a base for the 1-forms, such that $\Psi = \Psi_t(t,r) \, \extd t + \Psi_r(t,r) \, \extd r$. Hence, equation \eqref{8} can be expressed algebraically as a linear first-order PDE
\begin{equation}\label{psieq}
\partial_t \Psi_r - \partial_r \Psi_t = \frac{u'(r)}{4} \gamma_1 \Psi_t\,.
\end{equation}
It can be solved by separating in variables, $\Psi_t=T_t(t) R_t(r)$ and $\Psi_r=T_r(t) R_r(r)$ with the constraint $T_t =\partial_t T_r$ to solve the homogeneous sector in time. Next, by applying a diffeomorphism of the type $\extd t \rightarrow \extd t' = T_t(t) \extd t$, we absorb the time dependence of $\Psi_t$, and $\Psi_r$ must be at least linear on $t$. Thus, without any loss of generality, $\Psi = \Psi_t(r) \, \extd t + \big(\alpha(r) \, t + \beta(r) \big) \, \extd r$. But this expression can be, again, simplified by a gauge fixing of the type $t \rightarrow t'= t \, f(r) + g(r)$ with a suitable choice for the functions $f$ and $g$ in terms of $\Psi_t$, $\alpha$ and $\beta$. We finally obtain (after dropping the primes on $t$ and by redefining the function components)
\begin{equation}
    \Psi = \Psi(r) \,  \extd t\,. 
\end{equation}
We keep fixed our choice for the $t$ coordinate henceforth. Because of this gauge choice, the equations simplify enormously due to the vanishing of the fermion bilinears of the kind $\sim \bar{\Psi} \wedge \gamma \Psi$ (for any $\gamma$-matrix) and  $\Psi(r)$ can be solved from \eqref{psieq} given a certain prepotential function. 

Next, we turn to equation \eqref{4}
and plug in our previous expressions to yield an explicit solution for $\chi^{\alpha}$,
\begin{equation}\label{chisol}
\chi^{\alpha}(t,r) = t \, \Big( -\frac{u(X)}{2} (\gamma_\ast)^{\alpha \beta} \pm \sqrt{2 w(X) - 2M} (\gamma^0)^{\alpha \beta} \Big) \Psi_{\beta}(r)\,.
\end{equation}
From this solution, it is easy to see that any term of the kind $\bar{\chi} \Psi$ or $\bar{\chi} \gamma_1 \Psi$ will also vanish due to the antisymmetric contraction $\bar{\Psi} \gamma \Psi = 0$ of the spinorial fields. We are now able to use equation \eqref{3} and fix the Carroll boosts such that $\XP = 0$ (which is always possible locally) to obtain a constraint
\begin{equation}
\omega = - \frac{V}{\XH} \tau
\end{equation}
that causes the remaining equations for Carroll curvature \eqref{5} and for torsion \eqref{6} to be identical to each other. 
Plugging the constraint into the torsion equation simplifies it to
\begin{equation}
\extd \tau + ( \partial_X \, \text{ln} \XH) \tau \wedge \extd X = 0
\end{equation}
which is solved by
\begin{equation}
\tau = - \XH \, \extd t
\end{equation}
where we have fixed the residual Carroll boost invariance by assuming $\tau$ has no $\extd r$-component. This implies 
\begin{equation}
\omega = V(X) \, \extd t
\end{equation}
for the Carroll boost connection. Taking the exterior derivative of the former expression, it yields the same result for the Carrollian curvature as \eqref{5}, which is given by
\begin{equation}
\Omega = - V'(X) \, \tau \wedge e = - V'(X) \, \extd t \wedge \extd X\,.
\end{equation}
We can finally translate these results into the second-order formalism of the previous Section, where now the spatial metric yields
\begin{equation}
\extd s^2= e_{\mu} e_{\nu} \extd x^{\mu} \extd x^{\nu} = \extd r^2
\end{equation}
and the timelike vector field
\begin{equation}
v = v^{\mu} \partial_{\mu} = \frac{1}{\XH} \partial_t = \pm \frac{1}{\sqrt{2 w(X) - 2M}} \partial_t\,.
\end{equation}

The general deformed dilaton case ($Z\neq 0$) can be generated from the former solutions by applying a conformal transformation of the type \eqref{Weyl}. As equation \eqref{7} still holds, we keep fixed the radial coordinate by $e = \extd r$. After the Weyl transformation \eqref{Weyl}, the function $w(X)$ needs to be redefined as
\begin{equation}
w(X) := \int^X e^{-2Q(x)} V(x) \extd x \qquad \text{with} \qquad Q(X) = -\frac{1}{2} \int^X Z(x) \extd x
\end{equation}
where now the bosonic potential is given in terms of the prepotential according to \eqref{potential}. The additive integration constant of $w(X)$ can be absorbed in the definition of the conserved mass $M$, while the multiplicative one can be chosen to fix the desired physical dimensions of $V(X)$ (see \cite{Ecker:2023uwm} for a detailed discussion about choosing physical dimensions for this theories). The new Carroll mass and dilaton equations yields
\begin{equation}
M = w(X) - \frac{1}{2} \XH^2 e^{-2Q(X)} \qquad \qquad \extd r = \mp \frac{e^{-Q(X)} \, \extd X}{\sqrt{2 \,(w(X) - M)}}\,.
\end{equation}
For the fermionic sector we have that $\Psi(r) \rightarrow e^{Q(X)/2} \Psi(r)$, $\chi \rightarrow e^{-Q(X)/2} \chi$, hence the solution given by \eqref{chisol} still holds. The explicit form of $\Psi$, 
\begin{equation}
 \Psi = \Psi(X) \, \extd t =  e^{-Q(X)/2} \cdot \text{exp} \Big\lbrace \gamma_1 \, \int^X \frac{(u'(x) + \frac{u(x) Z(x)}{2})}{4 \XH(x)} \extd x \Big\rbrace \, \Psi_0 \, \extd t
\end{equation}
allows for a closed expression of $\chi(r,t)$ and the fermionic parnter of the curvature,
\begin{equation}
\sigma = \Big( \frac{V}{u} \gamma_1 + \frac{Z}{4} \XH \Big) \Psi(X) \, \extd t \wedge \extd r\,.
\end{equation}

We can proceed again as in the prepotential case and fix the Carroll boosts to demand $\XP=0$. The time einbein and spin connection are given by
\begin{equation}
\tau = - e^{-2Q(X)} \XH \, \extd t \qquad \qquad \omega =  e^{-2Q(X)} \big( V(X) - \frac{\XH^2}{2} Z(X) \big) \, \extd t
\end{equation}
and the Carroll curvature reads
\begin{equation}
\Omega = - \big( V'(X) - \frac{\XH^2}{2} Z^\prime(X) \big) e^{-2Q(X)} \, \extd t \wedge \extd X\,.
\end{equation}
An interesting aspect of this solution is that the Carroll curvature does not depend on the gravitino. In fact, the bosonic sector solution space is the same as in the pure bosonic theory. This is principally due to the vanishing of the soul contribution $\sim \bar{\chi}\chi$ of the bosonic potential when performing the Carroll contraction, which exhibits a clearly distinct feature from the Lorentzian case.

\subsubsection{Selected models}
Here, we present the Carroll $\mathcal{N}=1$ supersymmetric extensions of certain 2d dilaton models that are common in the literature.\footnote{%
In the Lorentzian version, these models are relevant because they serve as toy models for applications on 2d gravity holography, lower dimensional description of black holes, string theory, and black hole evaporation, among other phenomena.} 
In Table \ref{table:1}, we include a list of the generating potentials and derived functions of these supersymmetric models: JT gravity (sCJT), spherically reduced Schwarzschild model (sCS), CGHS model (sCCGHS), Cangemi--Jackiw model (sCCJ), which can also be obtained as a Weyl rescaled CGHS, and the $ab$-family (sCab), that includes all the previous ones by a suitable choice of the parameters $a$,$b$ and $B$. As mentioned previously, the generating prepotential is defined up to a sign, so the models are equivalent under the change $u \rightarrow -u$.
\begin{table}[ht!]
\centering
\begin{tabular}{||c|| c c | c c | c ||}
 \hline
 Model & $u(X)$ & $Z(X)$ & $V(X)$ & $w(X)$ & Reality condition\\ [0.5ex]
 \hline\hline
 \text{sCJT} & $-\frac{2X}{\ell}$ & $0$ & $\frac{X}{\ell ^2}$ & $\frac{X^2}{2\ell ^2}$ & $\ell\in\mathbb{R}$\\[.5em]
 \text{sCS} & $ 2 \lambda \sqrt{X}$ &$-\frac{1}{2X}$ & $\frac{\lambda ^2}{4}$ & $\frac{\lambda ^2}{2}\sqrt{X}$ & $\lambda\in\mathbb{R}, X\geq 0$\\[.5em]
 \text{sCCGHS} & $8 \lambda X$ &$-\frac{1}{X}$ & $2\lambda^2 X$ & $2\lambda^2 X$ & $\lambda\in\mathbb{R}$\\[.5em]
  \text{sCCJ} & $ \sqrt{8 \Lambda X}$ &$0$ & $\Lambda$ & $\Lambda X$ & $\Lambda X\geq 0$\\[.5em]
 \text{sCab} & $2 \sqrt{\frac{B}{b+1}} X^{\frac{a+b+1}{2}}$ &$-\frac{a}{X}$ & $\frac{B}{2}X^{a+b}$ & $\frac{B}{2(b+1)}X^{b+1}$ & $\frac{B}{b+1}\geq 0$, $X>0$\\[.5em]
 \hline
\end{tabular}
\caption{Selected Carroll dilaton SUGRA models, their potentials, and reality conditions.}
\label{table:1}
\end{table}

Extending these models to their supersymmetric Carrollian version limits the possible bosonic potentials by imposing certain convexity conditions on the dilaton dependence. The last column in Table \ref{table:1} makes this explicit by displaying restrictions from demanding the prepotential $u$ to be real. For example, the sCCJ model requires a non-negative coupling constant $\Lambda$ for the usual choice of non-negative dilaton $X$.

\subsection{Boundary conditions for sCJT}\label{se:2.4}
In this Section, we discuss a possible choice of boundary conditions for the sCJT model, where we show that the fermionic sector of the theory enhances the boundary charges by a non-trivial soul contribution. We start by changing the $H,B$ and $P$ basis to $L_{+},L_{-}$ and $L_{0}$ basis on the sCJT algebra \ref{calgebra}
\begin{equation}\label{basis}
L_{\pm} = \frac{1}{\ell} B \, \pm \, H  \qquad\qquad L_0 = P
\end{equation}
where the commutation relations given by
\begin{gather}
    \qty[ L_{\pm},L_0]= \pm \frac{1}{\ell} \, L_{\pm}
\end{gather}
show the existence of two subalgebras spanned by $\{ L_+, P \}$ and $\{ L_{-}, P \}$ respectively. The next step is to define an analogous change of basis for the supercharges by splitting the $Q_{\alpha} = (Q_1,Q_2)$ spinor into its two fermionic components. Following the conventions specified in Appendix \ref{Appendix:Notation}, we pick a basis where $\gamma_1$ is diagonal and get
\begin{equation}
Q_{\pm} = \frac{1}{\sqrt{2}} \big( Q_1 \mp Q_2 \big)\,.
\end{equation}
The non-vanishing (anti)commutation relations read
\begin{equation}
\big[ Q_{\pm}, L_0 \big] = \pm \frac{1}{2 \ell} Q_{\pm} \qquad\qquad \big\{ Q_+, Q_+ \big\} = - L_+ \qquad\qquad \big\{ Q_-, Q_- \big\} =  L_-\,.
\end{equation}

We introduce a coordinate system $(r,\tau)$ with $\tau$ being the coordinate along the boundary, which is located at $r \rightarrow \infty$. To specify boundary conditions for the scalars $\X$ and the gauge field $A$ we utilize the Coussaert--Henneaux--van Driel gauge \cite{Coussaert:1995zp} that splits the radial and temporal dependence  
\begin{equation}
    A =b^{-1} \big(\extd+a(\tau) \big)b \qquad\qquad \X=b^{-1} x(\tau) b
\end{equation}
and the boundary conditions in the highest-weight for $a = a_{\tau} \, \extd t$ are given by
\begin{equation}
a(\tau) = \big[ L_{+}+ \Lm (\tau) L_{-} + Q_+ + G(\tau) Q_- \big] \, \extd\tau\qquad \qquad b=\exp( r \, L_{0})\,.
\end{equation}
The dynamical field $\Lm$ is an arbitrary function of time $\tau$, while $G$ correspond to an arbitrary Grassmann-valued function of $\tau$. The gauge field yields
\begin{align}\label{at}
    A= P \, \extd r+(e^{r/\ell}\, L_{+}+ \Lm(\tau) \, e^{-r/\ell} \, L_{-} + e^{r/2\ell} \, Q_+ + G(\tau) \, e^{-r/2\ell} \, Q_-) \, \extd\tau\, .
\end{align}

The EOM require the 0-forms to be stabilizers of the auxiliary connection \eqref{at}, 
\begin{equation}
\extd x + \big[ a, x \big] = 0\,.
\end{equation}
The radial equation is trivially solved by the fact that both $x$ and $a$ only depend on time. After the decomposition
\begin{equation}
x(\tau) = X_+(\tau) L_+ + X_-(\tau) L_- + X_0(\tau) L_0 + \chi^+(\tau) Q_+ + \chi^-(\tau) Q_-
\end{equation}
the temporal one yields first-order equations for the 0-forms
\begin{align}
&\dot{X}_+ + \frac{X_0}{\ell} + \chi^+ = 0 \qquad \quad \dot{X}_- - \Lm \frac{X_0}{\ell} - G \, \chi^- = 0\\
\nonumber
&\dot{X}_0 = 0 \qquad \quad \dot{\chi}^+ + \frac{X_0}{2 \ell}=0 \qquad \quad  \dot{\chi}^- - \frac{G \, X_0}{2 \ell}=0 \,.
\end{align}
This system can be solved by demanding the fields $X_+, \chi^+$ not to grow linearly on time and integrating the function $g(\tau) = \int_0^{\tau} G(\tau') \,\extd\tau'$. Then, the 0-form defined in the dual basis yields
\begin{equation}
\X^\ast= X_+ \, e^{r/\ell}\, L_+^\ast + \big(X_- + g(\tau) \chi^- \big) e^{-r/\ell} \, L_-^\ast + \chi^- \, e^{-r/2 \ell} \, Q_-^\ast
\end{equation}
where now the functions whose temporal dependence is not explicitly stated are taken as integrating constants. The EOM given in Section \ref{eoms} are solved by these field configuration and \ref{at} up to subleading terms. Equivalently, using the dual basis definition
\begin{equation}
L^\ast_{\pm} = \ell \, B^\ast \mp H^\ast 
\end{equation}
we can go back to the original basis and compute the asymptotic expansion of the scalar fields as
\begin{align}\nonumber
&X(r,\tau) = \ell X_+  \, e^{r/\ell} + \ell \big(X_- + g(\tau) \chi^- \big) e^{-r/\ell}\\
&\XH(r,\tau) = -X_+ \, e^{r/\ell} + \big(X_- + g(\tau) \chi^- \big) e^{-r/\ell}\\
\nonumber
&\XP(r,\tau) = 0 \qquad \chi^+(r,\tau) = 0 \qquad \chi^-(r,\tau) = \chi^- e^{-r/2 \ell}\,.
\end{align}
This asymptotic form is preserved by the joint gauge transformations \ref{cboosts}, \ref{lambdaH} by taking $\lambda_P = \epsilon_+ = \epsilon_-=0$ together with
\begin{equation}
\lambda = \eta X_+ e^{r/\ell} + \eta  \big(X_- + g(\tau) \chi^- \big) e^{-r/\ell} \qquad \lambda_H = \ell \, \eta X_+ e^{r/\ell} - \ell \, \eta  \big(X_- + g(\tau) \chi^- \big) e^{-r/\ell} \,.
\end{equation}
Here, $\eta$ stands as a free parameter that realizes the transformation. To compute the boundary charges we work within the Brown--Henneaux prescription where the leading terms of the expansion are fixed, and the subleading terms contain state-dependent information. In that sense, we take $X_+ = \frac{1}{2 \ell}$ to be fixed, then consider $\delta X_- \neq 0$, $\delta \chi^- \neq 0$ and assume a slicing of the phase space where $\eta$ is state-independent. Furthermore, making the comparison with the pure bosonic case studied in \cite{Ecker:2023uwm}, we can identify $X_- = \ell \, M$ as the conserved Casimir mass \ref{Cmass}. The time-dependant term $\sim \chi^-$ is a soul contribution arising from this particular choice of boundary conditions. The variation of the boundary charges
\begin{equation}
\delta \mathcal{Q}[\lambda_I] = \frac{\kappa}{2 \pi} \big( \delta X \, \lambda + \delta \XH \, \lambda_H + \delta \XP \, \lambda_P + \delta \chi^+ \, \epsilon_+ + \delta \chi^- \, \epsilon_- \big)
\end{equation}
can be integrated in field space to an infinite tower of boundary charges given by
\begin{equation}
\mathcal{Q} = \frac{\kappa}{2 \pi} \eta \, \big( \ell M + g(\tau) \chi^- \big) \,.
\end{equation}

The bosonic term proportional to $M$ recovers the known result that the boundary charge is the Casimir, see, e.g., \cite{Grumiller:2021cwg} and Refs.~therein. The fermionic term proportional to the function $g(\tau)$ at first glance produces infinitely many modes, since $g$ is an arbitrary function. However, as suggested in \cite{Cadoni:1999ja} it is natural to consider charges averaged over time, especially in a Euclidean context where Euclidean time is periodic. Since $\chi^-$ only has a zero-mode contribution, all the modes of the function $g(\tau)$ apart from its zero-mode would be irrelevant. Thus, in such a scenario the fermionic tower collapses to a single fermionic charge.

While it would be rewarding to extend and generalize our analysis to other boundary conditions, other models, and construct Schwarzian-type of boundary actions, we leave such discussions for future work and move on to $\mathcal{N}=2$ Carroll dilaton SUGRA in the next Section.

\section{\texorpdfstring{$\boldsymbol{\mathcal{N}=2}$}{N=2} Carroll dilaton supergravity}\label{se:3}
\subsection{Ultra-relativistic JT supergravity}
In this Section, we construct the $\mathcal{N} = 2$ CJT supergravity introducing the non-vanishing commutation relations of the $\mathcal{N} = 2$ AdS$_2$ algebra as a starting point,
\begin{align}
\big[ K, P_0 \big] &=  P_{1} & \big[ K, P_1 \big] &=  P_{0} & \big[ P_0, P_1 \big] &= \frac{1}{\ell^2} K \nonumber\\\label{N=2alg}
\big[ K, Q_{\alpha}^i \big] &=  \frac{1}{2} (\gamma_{\ast} Q^i)_{\alpha} & \big[ P_{a}, Q_{\alpha}^i \big] &= \frac{1}{2 \ell} (\gamma_{a} Q^i)_{\alpha} & [U,Q_{\alpha}^i] &= -\frac{1}{2} \epsilon^{ij} Q_{\alpha}^j
\end{align}
\begin{equation*}
\lbrace Q_{\alpha}^i, Q_{\beta}^j \rbrace = \delta^{ij} \big( \gamma^{a} \big)_{\alpha \beta} P_{a} + \frac{\delta^{ij}}{\ell} \big( \gamma_\ast \big)_{\alpha \beta} K - \frac{\epsilon^{ij}}{\ell} \epsilon_{\alpha \beta} U
\end{equation*}
where the indices $i,j=1,2$ label the sets of the supercharges $Q^i$, $U$ is introduced as the bosonic generator of the R-symmetry and the antisymmetric symbol is taken to be $\epsilon^{12} = -\epsilon^{21} = 1$.

\subsubsection{Democratic scaling}\label{democratic}
To take the Carroll limit of the algebra, we first consider the democratic rescaling for the supercharges, 
\begin{equation}
H =  c \, P_0 \qquad\qquad B = c \, K \qquad\qquad P = P_1 \qquad\qquad \hat{U} = U \qquad\qquad \hat{Q}_{\alpha}^i = \sqrt{c} \, Q_{\alpha}^i
\end{equation}
In this case, as we treat both supercharge generators on equal footing, we call it a ``democratic'' scaling, adopting the nomenclature of Merbis and Lodato \cite{Lodato:2016alv}. Taking the $c \rightarrow $ limit and dropping the hats in the generators, we get the $\mathcal{N}=2$ democratic Carrollian AdS$_2$ algebra,
\begin{align}
\big[ B, P\big] &= H & \big[ H, P \big] &= \frac{1}{\ell^2} B  & \big[ P, Q_{\alpha}^i \big] &= \frac{1}{2 \ell} (\gamma_{1} Q^i)_{\alpha}\\\nonumber
\big[U,Q_{\alpha}^1\big] &= -\frac{1}{2} Q_{\alpha}^2 & \big[U,Q_{\alpha}^2\big] &= \frac{1}{2} Q_{\alpha}^1 &
\lbrace Q_{\alpha}^1, Q_{\beta}^1 \rbrace &= \lbrace Q_{\alpha}^2, Q_{\beta}^2 \rbrace = \big( \gamma^{0} \big)_{\alpha \beta} H + \frac{1}{\ell} \big( \gamma_{\ast} \big)_{\alpha \beta} B\,.
\end{align}
We construct the JT action as a BF theory the same way we did in the previous Section. Defining the gauge fields 1-forms and scalars
\begin{align}
A &= \tau \, H + e \, P + \omega \, B + T\, U + \bar{Q}^i \, \Psi^i\\
\nonumber
\mathcal{X}^\ast &= \XH \, H^\ast + \XP \, P^\ast + X\, B^\ast + \Pi \, U^\ast +\bar{\chi}^i \, Q^{\ast i}
\end{align}
we construct the covariant curvatures and then the action
\begin{multline}\label{demJT}
S_{\textrm{\tiny sJT}}^{\textrm{\tiny dem}} = \frac{\kappa}{2 \pi} \int_{M} \Big[X \big( \extd \omega + \frac{1}{\ell^2} \tau \wedge e + \frac{1}{2\ell} \bar{\Psi}^i \gamma_\ast \wedge \Psi^i \big) + \XH \big( \extd \tau + \omega \wedge e + \frac{1}{2} \bar{\Psi}^i \gamma^0 \wedge \Psi^i \big) \\
+ \XP \extd e + \Pi \, \extd T + \bar{\chi}^i \, \big( \extd \Psi^i - \frac{1}{2 \ell} e \wedge \gamma_ 1 \Psi^i + \frac{\epsilon^{ij}}{2} \, T \wedge \Psi^j \big)\Big]\,.
\end{multline}
This action results in two independent copies of the $\mathcal{N} =1$ case by eliminating the R-symmetry fields $T$ and $\Pi$.

\subsubsection{Post-Carrollian despotic scaling}
To construct the post-Carrollian version of this theory, we consider a central extension to the $\mathcal{N}=2$ algebra \eqref{N=2alg} given by an expansion of $P_1$, $U$ and $Q$ in higher orders of the parameter $1/c$ in the Carrollian limit. We follow the construction done in \cite{Ravera:2022buz} where the authors also consider a despotic rescaling for the supercharges. We start by defining 
\begin{equation}
Q^{\pm} = \frac{1}{\sqrt{2}} \big( Q^1 \pm i\, \gamma_1 Q^2 \big)
\end{equation}
where, redefining $U \rightarrow i\, U$, the decomposed algebra \eqref{N=2alg} takes the form
\begin{subequations}
\begin{align}
\big[ P_0, P_1 \big] &= \frac{1}{\ell^2} K & \big[ K, Q^{\pm} \big] &=  \frac{1}{2} \gamma_{\ast} Q^{\mp} & \big[ U, Q_{\alpha}^{\pm} \big] &= \pm \frac{1}{2} (\gamma_1 Q^{\pm})_{\alpha} \\
\big[ K, P_0 \big] &=  P_{1} & \big[ P_{0}, Q_{\alpha}^{\pm} \big] &= -\frac{1}{2 \ell} (\gamma^{0} Q^{\mp})_{\alpha} &  \lbrace Q_{\alpha}^{\pm}, Q_{\beta}^{\pm} \rbrace &= \big( \gamma_{1} \big)_{\alpha \beta} \big( P_1 \mp \frac{1}{\ell} U \big) \\
\big[ K, P_1 \big] &=  P_{0} &
[P_1,Q_{\alpha}^{\pm}] &= \frac{1}{2 \ell} (\gamma_1 Q^{\pm})_{\alpha} & \lbrace Q_{\alpha}^+, Q_{\beta}^- \rbrace &= \big( \gamma^{0} \big)_{\alpha \beta} P_{0} + \frac{1}{\ell} \big( \gamma_{\ast} \big)_{\alpha \beta} K \,.
\end{align}
\end{subequations}
The despotic Post-Carrollian contraction is performed by
\begin{subequations}
    \label{pcscaling}
\begin{align}
P_0 &= \frac{1}{c} H &P_1 &= P + \frac{1}{c^2} M & K&= \frac{1}{c} B\\
U &= U_1 + \frac{1}{c^2} U_2 & Q_{\alpha}^+ &= \hat{Q}_{\alpha}^+ + \frac{1}{c^2} R_{\alpha} & Q_{\alpha}^- &= \frac{1}{c} \, \hat{Q}_{\alpha}^-
\end{align}
\end{subequations}
where $M$, $U_2$, and $R$ are introduced as central extensions of translations, R-symmetry, and supercharges respectively. This contraction is called ``despotic'' since both supercharges $Q^+$ and $Q^-$ scale with different powers of $c$. Taking the limit $c \rightarrow 0$ and dropping the hats to clear notation, we end up with the $\mathcal{N}=2$ supersymmetric extended AdS$_2$ Carroll algebra, 
\begin{align}
\label{eq:wtf1}
\big[ B, P\big] &= H & \big[ H, P \big] &= \frac{1}{\ell^2} B  & \big[B, H\big] &= M\\\nonumber
\big[B,Q_{\alpha}^+\big] &= \frac{1}{2} (\gamma_\ast Q^-)_{\alpha} &\big[B,Q_{\alpha}^-\big] &= \frac{1}{2} (\gamma_\ast R)_{\alpha} & \big[H,Q_{\alpha}^+\big] &= -\frac{1}{2 \ell} (\gamma^0 Q^-)_{\alpha}\\\nonumber
\big[H,Q_{\alpha}^-\big] &= -\frac{1}{2 \ell} (\gamma^0 R)_{\alpha} &\big[P,Q_{\alpha}^{\pm}\big] &= \frac{1}{2 \ell} (\gamma_1 Q^{\pm})_{\alpha} &\big[P,R_{\alpha} \big] &= \frac{1}{2 \ell} (\gamma_1 R)_{\alpha}\\\nonumber
\big[M,Q_{\alpha}^+\big] &= \frac{1}{2 \ell} (\gamma_1 R)_{\alpha} &\big[U_1,Q_{\alpha}^{\pm}\big] &= \pm \frac{1}{2} (\gamma_1 Q_{\alpha}^{\pm}) & \big[U_1,R_{\alpha}\big] &= \big[U_2,Q_{\alpha}^+\big] = \frac{1}{2} (\gamma_1 R)_{\alpha}
\end{align}
along with
\begin{align}
\label{eq:wtf2}
\lbrace Q_{\alpha}^+, Q_{\beta}^+ \rbrace &= \big( \gamma_1 \big)_{\alpha \beta} P - \frac{1}{\ell} \big( \gamma_1 \big)_{\alpha \beta} U_1 & \lbrace Q_{\alpha}^+, Q_{\beta}^- \rbrace &=  \big( \gamma^0 \big)_{\alpha \beta} H + \frac{1}{\ell} \big( \gamma_{\ast} \big)_{\alpha \beta} B\\
\nonumber
\lbrace Q_{\alpha}^+, R_{\beta} \rbrace &= \big( \gamma_1 \big)_{\alpha \beta} M - \frac{1}{\ell} \big( \gamma_1 \big)_{\alpha \beta} U_2 & \lbrace Q_{\alpha}^-, Q_{\beta}^- \rbrace &= \big( \gamma_1 \big)_{\alpha \beta} M + \frac{1}{\ell} \big( \gamma_1 \big)_{\alpha \beta} U_2
\end{align}
the other (anti-)commutators being zero. 

A remarkable feature of the algebra \eqref{eq:wtf1}-\eqref{eq:wtf2} is that boosts no longer commute with the Hamiltonian, $[B,H]=M$, but rather commute into something that commutes with all bosonic generators.\footnote{This property is shared by bosonic post-Carrollian theories. We thank Florian Ecker and Patricio Salgado-Rebolledo for discussions about post-Carrollian theories.}

We can construct the JT action as we did before by defining 
\begin{align}
A &= \tau \, H + e \, P + \omega \, B + m \, M + T_1\, U_1 + T_2\, U_2 + \bar{Q}^+ \, \Psi^+ + \bar{Q}^- \, \Psi^- + \bar{R} \, \rho\\
\nonumber
\mathcal{X}^\ast &= \XH \, H^\ast + \XP \, P^\ast + X\, B^\ast + X_M \, M^\ast +\Pi_1 \, U_1^\ast +\Pi_2 \, U_2^\ast +\bar{\chi}^+ \, Q^{\ast+} +\bar{\chi}^- \, Q^{\ast-} + \bar{\lambda} \, R^\ast
\end{align}
yielding the action 
\begin{align}\label{JTdesp}
S_{\textrm{\tiny sJT}}^{\textrm{\tiny des}} =&\frac{\kappa}{2 \pi} \int_{M} \Big[X \big( \extd \omega + \frac{1}{\ell^2} \tau \wedge e + \frac{1}{\ell} \bar{\Psi}^+ \wedge \gamma_\ast \Psi^- \big) + \XH \big( \extd \tau + \omega \wedge e + \bar{\Psi}^+ \wedge \gamma^0 \Psi^- \big)\\\nonumber
&+\XP \big( \extd e + \frac{1}{2} \bar{\Psi}^+ \wedge \gamma_1 \Psi^+ \big) + X_M \big( \extd m + \omega \wedge \tau + \bar{\Psi}^+ \wedge \gamma_1 \rho + \frac{1}{2} \bar{\Psi}^- \wedge \gamma_1 \Psi^- \big)\\
\nonumber
&+\Pi_1 \big( \extd T_1 - \frac{1}{2 \ell} \bar{\Psi}^+ \wedge \gamma_1 \Psi^+ \big) + \Pi_2 \big( \extd T_2 - \frac{1}{2 \ell} \bar{\Psi}^+ \wedge \gamma_1 \rho + \frac{1}{2 \ell} \bar{\Psi}^- \wedge \gamma_1 \Psi^- \big) \\\nonumber 
& + \bar{\chi}^+ \big( \extd \Psi^+  - \frac{1}{2 \ell} e \wedge \gamma_1 \Psi^+ - \frac{1}{2} T_1 \wedge \gamma_1 \Psi^+ \big) + \bar{\chi}^- \big( \extd \Psi^- - \frac{1}{2} \omega \wedge \gamma_\ast \Psi^+ \\\nonumber 
&\qquad + \frac{1}{2 \ell} \tau \wedge \gamma^0 \Psi^+ - \frac{1}{2 \ell} e \wedge \gamma_1 \Psi^- + \frac{1}{2} T_1 \wedge \gamma_1 \Psi^- \big) + \bar{\lambda} \big( \extd \rho - \frac{1}{2} \omega \wedge \gamma_\ast \Psi^- \\\nonumber 
&\qquad + \frac{1}{2 \ell} \tau \wedge \gamma^0 \Psi^+ - \frac{1}{2 \ell} m \wedge \gamma_1 \Psi^+ - \frac{1}{2 \ell} e \wedge \gamma_1 \rho - \frac{1}{2} T_1 \wedge \gamma_1 \rho - \frac{1}{2} T_2 \wedge \gamma_1 \Psi^+ \big)\big]\,.
\end{align}

\subsection{Democratic Carroll dilaton supergravity}\label{section}
We recall from \cite{Bergamin:2004sr} the most general 2d dilaton supergravity Lorentzian theory. Due to our gauge choice for the internal R-symmetry algebra in the previous Section, here we only consider the twisted-chiral gauging for the Lorentzian case. We present it, however, with slight differences to adapt it to our conventions (see Appendix \ref{Appendix:Notation}). The Lorentzian action 
\begin{align}\nonumber\label{LN=2dil}
&S_{\textrm{\tiny sDIL}}^{L} =\frac{\kappa}{2 \pi} \int_{M}  \Big[X \extd \omega + X^a \Big( \extd e_a -\epsilon_{ab} \, \omega \wedge e^b + \frac{1}{2} \bar{\Psi}^i \, \gamma_a \wedge \Psi^i \Big) + \Pi \extd T \\
\nonumber
&+ \bar{\chi}^i \Big(\extd \Psi^i - \frac{1}{2} \omega \wedge \gamma_\ast \Psi^i + \frac{1}{2} T \wedge \Psi^j \, \epsilon^{ij} + \frac{1}{4} \text{Re}(W) \gamma^a \, e_a \wedge \Psi^i - \frac{1}{4} \text{Im}(W) \, e_a \wedge \gamma^a  \gamma_\ast \Psi^j \, \epsilon^{ij} \Big)\\
\nonumber
&+\frac{1}{4} \text{Re}(Z) X^b \bar{\chi}^i \Big( -\epsilon_b^a e_a \wedge \Psi^i + e_b \wedge \gamma_\ast \Psi^i \Big) + \frac{1}{4} \text{Im}(Z) X^b \bar{\chi}^i \Big(e_b \wedge \Psi^j - \epsilon_b^a e_a \wedge \gamma_\ast \Psi^j \Big) \epsilon^{ij}\\
&+ \big(V + \frac{X^c X_c}{2} \, \text{Re}(Z) \big) \frac{\epsilon_{ab}}{2} \, e^a \wedge e^b -\frac{1}{2} \text{Re}(U) \, \bar{\Psi}^i \, \gamma_\ast \wedge \Psi^i +\frac{1}{2} \text{Im}(U) \, \bar{\Psi}^i  \wedge \Psi^j \, \epsilon^{ij}\Big]
\end{align}
involves the potential functions
\begin{equation}
W= u' + \frac{\bar{Z} \, u}{2} \qquad  \qquad U = \frac{u}{2} + \frac{\bar{Z}}{8} \big( \bar{\chi}^i \chi^i + i \, \epsilon^{ij} \bar{\chi}^i \gamma_\ast \chi^j \big)\,.
\end{equation}
The prepotential functions $u(\mathcal{X})$ and $Z(\mathcal{X})$ depend only on the complex coordinate $\mathcal{X} \equiv X + i \Pi$ and we denote their complex conjugate as $\bar{u}(\mathcal{\bar{X}})$ and $\bar{Z}(\mathcal{\bar{X}})$. They are related to the potential function $V(X,\Pi, \chi^i_{\alpha})$ through
\begin{equation}
V = \frac{1}{8} \Big( (u \, \bar{u})' + \text{Re}(Z) u \, \bar{u} \Big) - \frac{\bar{\chi}^i \chi^i}{8} \, \text{Re}(w) + \frac{ \epsilon^{ij} \bar{\chi}^i \gamma_\ast \chi^j}{8} \, \text{Im}(w)
\end{equation}
where we use the notation $f^\prime = \partial_X (f)$ for derivatives and have defined
\begin{equation}
w(\mathcal{X}) \equiv \frac{Z^2}{4} \bar{u} + Z \bar{u}' + \frac{Z'}{2} \bar{u} + \bar{u}''\,.
\end{equation}
One particular case is the Lorentzian $\mathcal{N}=2$ JT supergravity, which can be recovered by taking $u =  -\frac{2}{\ell} (X + i \, \Pi)$ and $Z=0$.

To obtain the democratic Carrollian contraction of this theory, we proceed the same way as before and redefine the gauge fields [in agreement with the rescaling done in Section \eqref{democratic}] and the corresponding scalar fields to preserve the kinetic terms of the fields at second order in $c$,
\begin{align}
\omega &\rightarrow c \, \omega & e^0 &\rightarrow c \, \tau & e^1 &\rightarrow e & \Psi^i &\rightarrow c^{1/2} \, \Psi^i & T &\rightarrow T\\
X &\rightarrow c \, X & X^0 &\rightarrow - c \, \XH & X^1 &\rightarrow c^2 \XP & \chi^i &\rightarrow c^{3/2} \chi^i & \Pi &\rightarrow c^2 \Pi\,.
\end{align}

Truncating the action at order $c^2$, we retain $\mathcal{V}$ as an arbitrary potential generator of our theory, so it must scale as $\mathcal{V} \rightarrow c \, \mathcal{V} + \mathcal{O}(c^2)$. To achieve this, we need to demand simultaneously $u \rightarrow c \, u$ and $Z \rightarrow \frac{1}{c} \, Z$ at first order in the contraction parameter $c$. Expanding to second order in $c$ yields 
\begin{multline}
u (\mathcal{X} ) = \sum_n \, a_n \big( X + i \, \Pi \big)^n \rightarrow \sum_n \, a_n c^n \big( X + i \, c \Pi \big)^n = \, c \, \sum_n \, \hat{a}_n \big( X^n + c \, i \Pi \, n X^{n-1} + \mathcal{O}(c^2) \big) \\
= \, c \, u(X) + c^2 \, i \Pi \, u'(X) + \mathcal{O}(c^3)
\end{multline}
with the rescaling $\hat{a}_n := c^{n-1} a_n$. Now $u$ is real valued and only depend on $X$. For the other potentials, we proceed the same way and retain the expressions up to next-to-leading order in $c$, which naturally splits the functions into their real and imaginary components:
\begin{align}\label{rescale}
 u( \mathcal{X}) &\rightarrow c \, u(X) + c^2 \, i \Pi \, u'(X) + \mathcal{O}(c^3) &Z(\mathcal{X}) &\rightarrow \frac{1}{c} \, Z(X) + i \Pi \, Z'(X) + \mathcal{O}(c)\\
\nonumber
U( \mathcal{X}) &\rightarrow c  \frac{u(X)}{2} + \mathcal{O}(c^2) & W( \mathcal{X}) &\rightarrow W(X)  + \mathcal{O}(c)\,.
\end{align}
The Carrollian limit of \eqref{LN=2dil} (requiring expansion to second order in $c$) is
\begin{align}\label{demdil}
&S_{\textrm{\tiny sDIL}}^{\textrm{\tiny dem}} = \frac{\kappa}{2 \pi} \int_{M} \Big[X \extd \omega + \XH \big(\extd \tau + \omega \wedge e + \frac{1}{2} \bar{\Psi}^i \gamma^0 \wedge \Psi^i \big) + \XP \, \extd e  + \Pi \, \extd T \\
\nonumber
& + \bar{\chi}^i \, \big( \extd \Psi^i + \frac{\epsilon^{ij}}{2} \, T \wedge \Psi^j + \frac{V}{u}  e \wedge \gamma_ 1 \Psi^i + \frac{1}{4} Z \, \XH \, e \wedge \Psi^i \big)  + \big(V - \frac{\XH^2}{2} \, Z \big) \tau \wedge e   - \frac{u}{4} \bar{\Psi}^i \gamma_\ast \wedge \Psi^i\Big]
\end{align}
with the potential
\begin{equation}
V(X) = \frac{1}{8} \big( (u^2)' + Z \, u^2 \big)\,. 
\end{equation}

The democratic action \eqref{demdil} satisfies the graded Jacobi identities and can be thought of as two independent copies of the $\mathcal{N}=1$ case with a mixing term for the supercharges due to the R-symmetry. Notice that all the potentials depend only on the dilaton field $X$. We recover the democratic JT case \eqref{demJT} with the choice $u = -\frac{2}{\ell} X$ and $Z=0$.

\subsection{Despotic Carroll dilaton supergravity}
In this Section, we consider the despotic limit of the Lorentzian action \eqref{LN=2dil}. We express it in terms of the despotic spinor components $\Psi^{\pm}$ by defining
\begin{equation}
\Psi^{\pm} = \frac{1}{\sqrt{2}} \big( \Psi^1 \pm i \, \gamma_1 \Psi^2 \big)\qquad \qquad \bar{\chi}^{\pm} = \frac{1}{\sqrt{2}} \big( \bar{\chi}^1 \mp i \, \bar{\chi}^2 \gamma_1 \big)
\end{equation}
Next, we imitate the scaling \eqref{pcscaling},
where now we neglect the quadratic post-Carrollian extended terms,
\begin{align}
\omega &\rightarrow c \, \omega & e^0 &\rightarrow c \, \tau & e^1 &\rightarrow e & \Psi^+ &\rightarrow \Psi^+ & \Psi^- &\rightarrow c \, \Psi^- & T &\rightarrow T\\
\nonumber
X &\rightarrow c \, X & X^0 &\rightarrow - c \, \XH & X^1 &\rightarrow c^2 \XP & \chi^+ &\rightarrow c^{2} \chi^+ & \chi^- &\rightarrow c \chi^- & \Pi &\rightarrow c^2 \Pi\,.
\end{align}

We start by considering the $Z=0$ case for simplicity. Then, we rescale the potential functions following the same procedure of the previous Section in \eqref{rescale} up to second order. Finally, by redefining $T \rightarrow -i \, T$ and $\Pi \rightarrow i \, \Pi$ to retain a real action and truncating at lowest order in $c$, we get the despotic action
\begin{align}\label{desdil}
&S_{\textrm{\tiny sDIL}}^{\textrm{\tiny des}} = \frac{\kappa}{2 \pi} \int_{M} \Big[X \extd \omega + \XH \big( \extd \tau + \omega \wedge e + \bar{\Psi}^+ \wedge \gamma^0 \Psi^- \big) + \XP \big( \extd e + \frac{1}{2} \bar{\Psi}^+ \wedge \gamma_1 \Psi^+ \big) \\\nonumber
&+ \Pi \extd T  + \bar{\chi}^+ \big( \extd \Psi^+ - \frac{1}{2} T \wedge \gamma_1 \Psi^+ + \frac{u'}{4} e \wedge \gamma_1 \Psi^+ \big) \\\nonumber
&+ \bar{\chi}^- \Big( \extd \Psi^- - \frac{1}{2} \omega \wedge \gamma_\ast \Psi^+ + \frac{1}{2} T \wedge \gamma_1 \Psi^- - \frac{u'}{4} \tau \wedge \gamma^0 \Psi^+ + \frac{u'}{4} e \wedge \gamma_1 \Psi^- + \frac{\Pi}{4} u'' e \wedge \gamma_\ast \Psi^+\Big)\\
\nonumber
& + \Big( V(X) - \frac{\bar{\chi}^- \chi^-}{8} u'' \Big) \, \tau \wedge e  + \Pi \, \frac{u'}{4} \bar{\Psi}^+ \wedge \gamma_1 \Psi^+ - \frac{u}{2} \bar{\Psi}^+ \wedge \gamma_\ast \Psi^-\Big]\,.
\end{align}
We denote the prepotential $u$ and its derivatives as real-valued and only function of the dilaton field $X$. We can see as an interesting feature that in this despotic limit we get a non-vanishing soul contribution to the bosonic potential and to the intrinsic torsion. As a consistency check, by taking the usual JT prepotential $u= -\frac{2}{\ell} X$ and without considering the post-Carrollian terms, we can recover \eqref{JTdesp}. 

To generate the general deformed case ($Z \neq 0$), we proceed as in Section \ref{generaldif}, apply the target space diffeomorphism 
\begin{equation}
X^I (\hat{X}^J) = \big(\hat{X}, e^{Q(\hat{X})} \hat{X}_H, e^{Q(\hat{X})} \hat{X}_P, \hat{\Pi}, e^{\frac{1}{2}Q(\hat{X})} \hat{\chi}^{+}, e^{\frac{1}{2}Q(\hat{X})} \hat{\chi}^{-} \big)
\end{equation}
and compute the Poisson tensor using the transformation rule \ref{poisdiff} with $Z= -2Q'$. The non-vanishing components are
\begin{align}
&P^{\XH \XP} = V - \frac{\XH^2}{2} Z - \frac{\bar{\chi}^- \chi^-}{8} N \qquad \qquad \quad P^{X \XP} = \XH\\
\nonumber
&P^{X \alpha^+} = - \frac{\bar{\chi}^-}{2} \gamma_\ast^{\alpha^+} \qquad \qquad \quad  P^{T \alpha^+} = - \frac{\bar{\chi}^+}{2} \gamma_1^{\alpha^+} \qquad  \qquad \quad P^{T \alpha^-} = \frac{\bar{\chi}^-}{2} \gamma_1^{\alpha^-}\\
\nonumber
&P^{\XH \alpha^+} = \bar{\chi}^- \Big( -\frac{W}{4} \gamma^{0\alpha^+} + \frac{Z}{4} \XH \gamma_\ast^{\alpha^+} \Big) \qquad \qquad P^{\XH \alpha^-} = \bar{\chi}^- \Big( \frac{W}{4} \gamma_1^{\alpha^-} + \frac{Z}{4} \XH \delta^{\alpha^-} \Big)\\
\nonumber
&P^{\XP \alpha^+} = \bar{\chi}^- \Big( \frac{\Pi}{4} \, N + \frac{Z}{4} \XP \Big) \gamma_\ast^{\alpha^+} + \bar{\chi}^+ \Big( \frac{W}{4} \gamma_1^{\alpha^+} + \frac{Z}{4} \XH \delta^{\alpha^+} \Big)\\
\nonumber
&P^{\alpha^+ \beta^+} = - \big(\XP + \frac{\Pi}{2} \, W \big) \big( \gamma_1 \big)^{\alpha^+ \beta^+} - \frac{Z}{8} \Big( \bar{\chi}^{+ \, \alpha^+} \bar{\chi}^- \gamma_\ast^{\beta^+} + \bar{\chi}^{+ \, \beta^+} \bar{\chi}^- \gamma_\ast^{\alpha^+} \Big)\\
&P^{\alpha^+ \alpha^-} = -\XH \big(\gamma^0 \big)^{\alpha^+ \alpha^-} + \Big(\frac{u}{2} + \frac{Z}{16} \bar{\chi}^- \chi^- \Big) \gamma_\ast^{\alpha^+ \alpha^-} \nonumber
\end{align}

where we established the convention that the fermionic indices of the kind $\alpha^+$ are contracted only with the $\psi^+$ spinors and the $\alpha^-$ with the $\psi^-$.The functions $V(X)$, $W(X)$ and $N(X)$ are given only in terms of the generating potentials $u(X)$ and $Z(X)$ by
\begin{equation}
W = u' + \frac{u \, Z}{2}\qquad \qquad V = \frac{u \, W}{4}  \qquad \qquad N= W' + \frac{W \, Z}{2} \,.
\end{equation}

\section{Conclusions}\label{se:4}

Having constructed generic $\mathcal{N}=1$ (Section \ref{se:2}) and $\mathcal{N}=2$ (Section \ref{se:3}) Carroll dilaton supergravity in two dimensions, we discuss now some future applications of the $\mathcal{N}=1$ action \eqref{csDIL1st}, the democratic $\mathcal{N}=2$ action \eqref{demdil}, and the despotic $\mathcal{N}=2$ action \eqref{desdil}, which at the same time serves as motivation for our work.

Holography is a key motivation. Having precise holographic correspondences available in relatively simple models, such as the SYK/JT correspondence \cite{Kitaev:15ur,Sachdev:1992fk,Sachdev:2010um, Maldacena:2016hyu,Mertens:2018fds,Sarosi:2017ykf,Gu:2019jub}, is useful for our conceptual understanding of quantum gravity and holography. In this context, lower-dimensional models play a key role due to their technical manageability. On the gravity side, the lowest meaningful dimension is two, assuming we want to have black holes as part of the spectrum; in this sense, two-dimensional gravity models are the most efficient ones in terms of technical simplicity (see \cite{Grumiller:2002nm} for a review).
Supersymmetry provides tools and properties that allow more analytic control, such as supersymmetric localization \cite{Pestun:2007rz} or simple proofs of positivity of gravitational energy in four \cite{Witten:1981mf} or two \cite{Park:1993sd} dimensions. Scaling limits can zoom into specific sectors of the original theory, provide a parameter for a perturbative expansion, lead to significant simplification, and exhibit new phenomena. All of the above applies to the Carrollian limit of dilaton supergravity in two dimensions studied in our work.

Aiming for a novel SYK/JT-like correspondence, we now address possible future steps. First and foremost, one should generalize our Section \ref{se:2.4} where we discussed an example of boundary conditions for sCJT. However, our analysis was not nearly as exhaustive as the bosonic counterpart (see \cite{Grumiller:2017qao}) and it should be rewarding to fill this gap. Moreover, other noteworthy Carroll dilaton supergravity models besides sCJT could be amenable to a holographic description, e.g., the sCCGHS or the sCCJ model, see Table \ref{table:1} for details.

Along the lines of \cite{Grumiller:2020elf}, one might be able to construct Schwarzian-type boundary actions for a given choice of boundary conditions. In the best-case scenario, such a boundary action can also be derived on the field theory side, using an exotic SYK-like model. Depending on the details, it may be possible to derive such a model as a scaling limit (mimicking the Carrollian limit on the gravity side of the holographic correspondence) from a more ordinary SYK-like model.

\section*{Acknowledgments}

This work was supported by the Austrian Science Fund (FWF), projects P~33789 and P~36619. We thank the Erwin-Schr\"odinger Institute (ESI) for the hospitality in April 2024 during the program \href{https://www.esi.ac.at/events/e518/}{``Carrollian physics and holography''} and we are grateful to the participants of this program for numerous insightful discussions. We thank particularly Ankit Aggarwal, Arjun Bagchi, Luca Ciambelli, Florian Ecker, Adrien Fiorucci, Jelle Hartong, Stefan Prohazka, Patricio Salgado-Rebolledo, Marc Henneaux, and Dima Vassilevich for discussions and collaborations on related topics.

MSN was supported by the Federal Ministry of Education, Science, and Research, OeAD-GmbH Agency for International Mobility and Cooperation in Education, Science, and Research. DG and LM acknowledge support from the OeAD travel grant IN 04/2022 and thank Rudranil Basu for hosting them at BITS Pilani in Goa in February 2024 through the grant DST/IC/Austria/P-9/202 (G), where LM presented some of the results of this work.

\appendix
\section{Notations and conventions} \label{Appendix:Notation}

\subsection{Index notation}\label{app:1}

We use lower case Latin characters from the beginning of the alphabet, $a,b,\dots$, for tangent space indices, Greek ones from the beginning of the alphabet, $\alpha,\beta,\dots$, for spinor indices, Greek ones from the middle of the alphabet, $\mu,\nu,\dots$, for spacetime indices, lower case Latin ones from the middle of the alphabet, $i,j,\dots$, for labeling the supercharges in the $\mathcal{N}=2$ case, and upper case Latin ones from the middle of the alphabet, $I,J,\dots$, for (co-)adjoint indices (in upper position they refer to co-adjoint objects, in lower case position to adjoint ones).

The supergravity conventions we use are in complete agreement with \cite{Freedman:2012zz}, so here we recall their most relevant aspects. To contract Grassman-valued fermonic quantities $\Psi$ and $\lambda$, we denote $\bar{\Psi} \lambda$, which in index notation can be written as $\bar{\Psi^{\alpha}} \lambda_{\alpha}$. For raising and lowering spinorial indices, we work in the NW-SE spinor convention and use the antisymmetric tensor
\begin{equation}
\epsilon^{\alpha \beta} = \epsilon_{\alpha \beta}=
\begin{pmatrix}
0 & 1\\
-1 & 0 
\end{pmatrix}
\end{equation}
such that $\epsilon^{01}=\epsilon_{01}=+1$ and $\epsilon^{\alpha \beta} \epsilon_{\sigma \beta} = \delta_{\sigma}^{\alpha}$. Any barred spinor is defined only with an upper index and non-barred with a lower index. Indices can be raised with $\bar{\Psi}^{\alpha} = \epsilon^{\alpha \sigma} \Psi_{\sigma}$ and lowered with $\Psi_{\alpha} = \bar{\Psi}^{\sigma} \epsilon_{\sigma \alpha}$. For practical reasons, we drop the bar while using index notation, so when expressing these quantities in compact form, $\chi$ refers to $\chi_{\alpha}$ and $\bar{\chi}$ for $\chi^{\alpha}$. Sometimes we also use abbreviations such as $(\gamma_a Q)_\alpha$, by which we mean
\eq{
(\gamma_a Q)_\alpha = (\gamma_a)_\alpha{}^\beta Q_\beta\,.
}{eq:appendix}

For raising or lowering spinorial indices on any $\gamma$-matrix, we have introduced the charge conjugation matrix $\mathcal{C}$ such that $\gamma_{\alpha}{}^{\beta} \rightarrow \big( \gamma \, \mathcal{C}^{-1} \big)_{\alpha \beta}$. In two dimensions, $\mathcal{C}^{-1}_{\alpha \beta} = \epsilon_{\alpha \beta}$, so it amounts to contract the antisymmetric symbol with an index of the matrix to upper/lower it. In the present paper, we drop the $\mathcal{C}$ factor on raised/lowered matrices to simplify notation. As an example, 
\begin{equation}
\bar{\Psi} \gamma \lambda = \Psi^{\alpha} \gamma_{\alpha}{}^{\beta} \lambda_{\beta}= - \Psi_{\alpha} \gamma^{\alpha \beta} \lambda_{\beta}
\end{equation}
where 
\begin{equation}\label{raising}
\gamma^{\alpha \beta} := \big( \mathcal{C} \, \gamma \big)^{\alpha \beta} = \epsilon^{\alpha \sigma} \gamma_{\sigma}{}^{\beta}\qquad \qquad \gamma_{\alpha \beta} := \big( \gamma \, \mathcal{C}^{-1} \big)_{\alpha \beta} = \gamma_{\alpha}{}^{\sigma} \epsilon_{\sigma \beta}\,.
\end{equation}

With our definition of the gamma matrices in two-dimensions, the spinorial indices convention and taking into account the fact that the spinors have Grassmann valued components that anticommute, the following relations can be proven:
\begin{equation}
\bar{\Psi} \lambda = \bar{\lambda} \Psi\qquad \qquad \bar{\Psi} \gamma \lambda =  - \bar{\lambda} \gamma \Psi\qquad \qquad (\overline{\gamma \Psi}) = - \bar{\Psi} \gamma
\end{equation}
When dealing with spinorial 1-forms $\Psi$ and $\lambda$, the anticommutativity of the wedge product inverts these relations,
\begin{equation}
\bar{\Psi} \wedge \lambda = - \bar{\lambda} \wedge \Psi\qquad \qquad \bar{\Psi} \wedge \gamma \lambda =  \bar{\lambda} \wedge \gamma \Psi\,.
\end{equation}

\subsection{Lorentzian conventions} \label{app:2}

The 2d Lorentzian tangent space metric is defined as $\eta_{ab} = \textrm{diag}(-1,1)$. For the tangent space antisymmetric contractions, we adopt the tensor $\epsilon_{ab}$ to be $\epsilon_{01} = -\epsilon_{10}=1$. The gamma matrices $\gamma^a =(\gamma^{0},\gamma^{1})$ satisfy the usual Lorentzian Clifford algebra $\lbrace \gamma^a, \gamma^b \rbrace = 2 \eta^{ab}$ and are given in the representation
\eq{
\big(\gamma^0 \big)_{\alpha}{}^{\beta} = - \big(\gamma_0 \big)_{\alpha}{}^{\beta} =
\begin{pmatrix}
0 & 1\\
-1 & 0
\end{pmatrix}
\quad\;\,
\big(\gamma^1 \big)_{\alpha}{}^{\beta} =  \big(\gamma_1 \big)_{\alpha}{}^{\beta} =
\begin{pmatrix}
0 & 1\\
1 & 0
\end{pmatrix}
\quad\;\,
 \big( \gamma_\ast \big)_{\alpha}{}^{\beta} =
\begin{pmatrix}
1 & 0\\
0 & -1
\end{pmatrix}
}{eq:app7}
where we have defined the gamma matrix featuring in the chiral projectors as 
\begin{equation}
\gamma_\ast = \epsilon_{ab} \gamma^a \gamma^b = \gamma^0 \gamma^1
\end{equation}

Raising the lower fermionic index according to \ref{raising}, we get the symmetric upper charge conjugated version of these matrices:
\eq{
\big(\gamma^0 \big)^{\alpha \beta} = -
\begin{pmatrix}
1 & 0\\
0 & 1
\end{pmatrix}
\qquad\;\,
\big(\gamma^1 \big)^{\alpha \beta} =
\begin{pmatrix}
1 & 0\\
0 & -1
\end{pmatrix}
\qquad\;\,
 \big( \gamma_\ast \big)^{\alpha \beta} = -
\begin{pmatrix}
0 & 1\\
1 & 0
\end{pmatrix}
}{eq:app7a}

\subsection{Carrollian conventions} \label{app:3}

While it turns out that we can work with the Lorentzian conventions for the gamma matrices even in the Carrollian case, we start here by discussing aspects unique to Carrollian spacetimes. For more details see, e.g., \cite{Banerjee:2022ocj,Bagchi:2022eui}.

The flat space Carroll metric $h_{ab}=\textrm{diag}(0,1)$ has no inverse, but we can nevertheless define a degenerate metric with upper indices, $h^{ab}=\textrm{diag}(-1,0)$. Note, however, that there is no device to raise or lower indices.

The Carroll--Clifford algebra
\eq{
2h_{ab}=\{\tilde\gamma_a,\,\tilde\gamma_b\}\qquad\qquad 2h^{ab}=\{\tilde\gamma^a,\,\tilde\gamma^b\}
}{eq:app1}
yields a nilpotent lower-index gamma matrix $\tilde\gamma_0^2=0$ and another one that squares to unity, $\tilde\gamma_1^2=\mathds{1}$. The versions with upper indices behave oppositely, $(\tilde\gamma^0)^2=-\mathds{1}$ and $(\tilde\gamma^1)^2=0$.

We choose the matrix representation\footnote{%
Apart from usual ambiguities from the choice of basis, the degenerate matrices $\tilde\gamma_0$ and $\tilde\gamma^1$ have scaling ambiguities, $\tilde\gamma_0\to\lambda_0\tilde\gamma_0$ and $\tilde\gamma_1\to\lambda_1\tilde\gamma_1$. We fix them such that $\{\tilde\gamma^0,\,\tilde\gamma_0\}=\{\tilde\gamma^1,\,\tilde\gamma_1\}=2\,\mathds{1}$.} 
\eq{
\tilde\gamma_0=\begin{pmatrix}
    -1 & -1 & \\ 1 & 1
\end{pmatrix} \qquad\quad
\tilde\gamma_1=\begin{pmatrix}
    0 & 1 \\ 1 & 0
\end{pmatrix} \qquad\quad
\tilde\gamma^0=\begin{pmatrix}
    0 & 1 & \\ -1 & 0
\end{pmatrix} \qquad\quad
\tilde\gamma^1=\begin{pmatrix}
    -i & 1 \\ 1 & i
\end{pmatrix}
}{eq:app2}
compatible with the identities $\tilde\gamma_0\tilde\gamma_1=-\tilde\gamma_1\tilde\gamma_0=\tilde\gamma_0$ and $\tilde\gamma^0\tilde\gamma^1=-\tilde\gamma^1\tilde\gamma^0=i\tilde\gamma^1$. Note that the non-degenerate gamma matrices, $\tilde\gamma^0$ and $\tilde\gamma_1$, are identical to their Lorentzian counter parts \eqref{eq:app7}. This is why in the main text we drop all tilde decorations, since we exclusively use $\tilde\gamma^0$, $\tilde\gamma_1$, and their product,
\eq{
\tilde\gamma_\ast := \tilde\gamma^0\,\tilde\gamma_1\,.
}{eq:app3}
Also $\tilde\gamma_\ast$ is identical to its Lorentzian counter part in \eqref{eq:app7} and hence again we drop all tildes in the main text. The gamma matrix $\tilde\gamma_\ast$ obeys the usual relations
\eq{
\{\tilde\gamma_\ast,\,\tilde\gamma^0\}=0=\{\tilde\gamma_\ast,\,\tilde\gamma_1\}\qquad\qquad\tilde\gamma_\ast^2=\mathds{1}\,.
}{eq:app4}

In terms of Pauli matrices, the Carroll-Clifford algebra representation above is given by $\tilde\gamma_1=\gamma_1=\sigma_1$, $\tilde\gamma^0=\gamma^0=i\sigma_2$, $\tilde\gamma_\ast=\gamma_\ast=\sigma_3$, $\tilde\gamma^1=-i\sigma_3+\sigma_1$, and $\tilde\gamma_0=-\sigma_3-i\sigma_2$. 


\providecommand{\href}[2]{#2}\begingroup\raggedright\endgroup

\begin{thebibliography}{10}

\bibitem{Park:1993sd}
Y.-C. Park and A.~Strominger, ``Supersymmetry and positive energy in classical and quantum two-dimensional dilaton gravity,'' {\em Phys. Rev.} {\bf D47} (1993) 1569--1575,
\href{http://www.arXiv.org/abs/arXiv:hep-th/9210017}{{\tt arXiv:hep-th/9210017}}.

\bibitem{Howe:1978ia}
P.~S. Howe, ``Super {W}eyl transformations in two-dimensions,'' {\em J. Phys.} {\bf A12} (1979)
393--402.

\bibitem{Freedman:1976xh}
D.~Z. Freedman, P.~van Nieuwenhuizen, and S.~Ferrara, ``{Progress Toward a Theory of Supergravity},'' {\em Phys. Rev. D} {\bf 13} (1976) 3214--3218.

\bibitem{Deser:1976eh}
S.~Deser and B.~Zumino, ``{Consistent Supergravity},'' {\em Phys. Lett. B} {\bf 62} (1976) 335.

\bibitem{Ikeda:1994dr}
N.~Ikeda, ``Gauge theory based on nonlinear {L}ie superalgebras and structure of 2-d dilaton supergravity,'' {\em Int. J. Mod. Phys.} {\bf A9} (1994)
1137--1152.

\bibitem{Izquierdo:1998hg}
J.~M. Izquierdo, ``Free differential algebras and generic 2d dilatonic (super)gravities,'' {\em Phys. Rev.} {\bf D59} (1999) 084017,
\href{http://www.arXiv.org/abs/arXiv:hep-th/9807007}{{\tt arXiv:hep-th/9807007}}.

\bibitem{Strobl:1999zz}
T.~Strobl, ``Target-superspace in 2d dilatonic supergravity,'' {\em Phys. Lett.} {\bf B460} (1999) 87--93,
\href{http://www.arXiv.org/abs/arXiv:hep-th/9906230}{{\tt arXiv:hep-th/9906230}}.

\bibitem{Ertl:2000si}
M.~Ertl, W.~Kummer, and T.~Strobl, ``General two-dimensional supergravity from {P}oisson superalgebras,'' {\em JHEP} {\bf 01} (2001) 042,
\href{http://www.arXiv.org/abs/arXiv:hep-th/0012219}{{\tt arXiv:hep-th/0012219}}.

\bibitem{Bergamin:2003mh}
L.~Bergamin, D.~Grumiller, and W.~Kummer, ``Supersymmetric black holes in 2d dilaton supergravity: baldness and extremality,'' {\em J. Phys.} {\bf A37} (2004) 3881--3901,
\href{http://www.arXiv.org/abs/hep-th/0310006}{{\tt hep-th/0310006}}.

\bibitem{LevyLeblond1965}
J.-M. L\'evy-Leblond, ``{Une nouvelle limite non-relativiste du groupe de Poincar\'e},'' {\em A. Inst. Henri Poincar\'e III 1} (1965).

\bibitem{Gupta1966}
N.~D. SenGupta, ``On an analogue of the galilei group,'' {\em Il Nuovo Cimento A Series 10} {\bf 44} (1966), no.~2, 512--517.

\bibitem{ESI2024}
A.~Campoleoni, L.~Donnay, S.~Fredenhagen, and D.~Grumiller, ``Carrollian physics and holography.'' {ESI Program} \href{https://www.esi.ac.at/events/e518/}{https://www.esi.ac.at/events/e518/}, 2024.

\bibitem{Ravera:2022buz}
L.~Ravera and U.~Zorba, ``{Carrollian and non-relativistic Jackiw\textendash{}Teitelboim supergravity},'' {\em Eur. Phys. J. C} {\bf 83} (2023), no.~2, 107, \href{http://www.arXiv.org/abs/2204.09643}{{\tt 2204.09643}}.

\bibitem{Ravera:2019ize}
L.~Ravera, ``{AdS Carroll Chern-Simons supergravity in 2 + 1 dimensions and its flat limit},'' {\em Phys. Lett. B} {\bf 795} (2019) 331--338, \href{http://www.arXiv.org/abs/1905.00766}{{\tt 1905.00766}}.

\bibitem{Ali:2019jjp}
F.~Ali and L.~Ravera, ``{$\mathcal{N}$-extended Chern-Simons Carrollian supergravities in $2+1$ spacetime dimensions},'' {\em JHEP} {\bf 02} (2020) 128, \href{http://www.arXiv.org/abs/1912.04172}{{\tt 1912.04172}}.

\bibitem{Bergshoeff:2015wma}
E.~Bergshoeff, J.~Gomis, and L.~Parra, ``{The Symmetries of the Carroll Superparticle},'' {\em J. Phys. A} {\bf 49} (2016), no.~18, 185402, \href{http://www.arXiv.org/abs/1503.06083}{{\tt 1503.06083}}.

\bibitem{Cardenas:2018krd}
M.~C\'ardenas, O.~Fuentealba, H.~A. Gonz\'alez, D.~Grumiller, C.~Valc\'arcel, and D.~Vassilevich, ``{Boundary theories for dilaton supergravity in 2D},'' {\em JHEP} {\bf 11} (2018) 077, \href{http://www.arXiv.org/abs/1809.07208}{{\tt 1809.07208}}.

\bibitem{Grumiller:2020elf}
D.~Grumiller, J.~Hartong, S.~Prohazka, and J.~Salzer, ``{Limits of JT gravity},'' {\em JHEP} {\bf 02} (2021) 134, \href{http://www.arXiv.org/abs/2011.13870}{{\tt 2011.13870}}.

\bibitem{Gomis:2020wxp}
J.~Gomis, D.~Hidalgo, and P.~Salgado-Rebolledo, ``{Non-relativistic and Carrollian limits of Jackiw-Teitelboim gravity},'' {\em JHEP} {\bf 05} (2021) 162, \href{http://www.arXiv.org/abs/2011.15053}{{\tt 2011.15053}}.

\bibitem{Hansen:2021fxi}
D.~Hansen, N.~A. Obers, G.~Oling, and B.~T. S\o{}gaard, ``{Carroll Expansion of General Relativity},'' {\em SciPost Phys.} {\bf 13} (2022), no.~3, 055, \href{http://www.arXiv.org/abs/2112.12684}{{\tt 2112.12684}}.

\bibitem{Campoleoni:2022ebj}
A.~Campoleoni, M.~Henneaux, S.~Pekar, A.~P\'erez, and P.~Salgado-Rebolledo, ``{Magnetic Carrollian gravity from the Carroll algebra},'' {\em JHEP} {\bf 09} (2022) 127, \href{http://www.arXiv.org/abs/2207.14167}{{\tt 2207.14167}}.

\bibitem{Ecker:2023uwm}
F.~Ecker, D.~Grumiller, J.~Hartong, A.~P\'erez, S.~Prohazka, and R.~Troncoso, ``{Carroll black holes},'' {\em SciPost Phys.} {\bf 15} (2023), no.~6, 245, \href{http://www.arXiv.org/abs/2308.10947}{{\tt 2308.10947}}.

\bibitem{Bergshoeff:2023vfd}
E.~A. Bergshoeff, A.~Campoleoni, A.~Fontanella, L.~Mele, and J.~Rosseel, ``{Carroll fermions},'' {\em SciPost Phys.} {\bf 16} (2024), no.~6, 153, \href{http://www.arXiv.org/abs/2312.00745}{{\tt 2312.00745}}.

\bibitem{Izawa:1999ib}
K.~I. Izawa, ``On nonlinear gauge theory from a deformation theory perspective,'' {\em Prog. Theor. Phys.} {\bf 103} (2000) 225--228,
\href{http://arXiv.org/abs/hep-th/9910133}{{\tt hep-th/9910133}}.

\bibitem{Ikeda:1993fh}
N.~Ikeda, ``{Two-dimensional gravity and nonlinear gauge theory},'' {\em Annals Phys.} {\bf 235} (1994) 435--464,
\href{http://www.arXiv.org/abs/hep-th/9312059}{{\tt hep-th/9312059}}.

\bibitem{Schaller:1994es}
P.~Schaller and T.~Strobl, ``Poisson structure induced (topological) field theories,'' {\em Mod. Phys. Lett.} {\bf A9} (1994) 3129--3136,
\href{http://arXiv.org/abs/hep-th/9405110}{{\tt hep-th/9405110}}.

\bibitem{Berends:1984rq}
F.~A. Berends, G.~J.~H. Burgers, and H.~van Dam, ``{On the Theoretical Problems in Constructing Interactions Involving Higher Spin Massless Particles},'' {\em Nucl. Phys. B} {\bf 260} (1985) 295--322.

\bibitem{Barnich:1993vg}
G.~Barnich and M.~Henneaux, ``Consistent couplings between fields with a gauge freedom and deformations of the master equation,'' {\em Phys. Lett.} {\bf B311} (1993) 123--129,
\href{http://arXiv.org/abs/hep-th/9304057}{{\tt hep-th/9304057}}.

\bibitem{Gomis:1995jp}
J.~Gomis and S.~Weinberg, ``{Are nonrenormalizable gauge theories renormalizable?},'' {\em Nucl. Phys. B} {\bf 469} (1996) 473--487, \href{http://www.arXiv.org/abs/hep-th/9510087}{{\tt hep-th/9510087}}.

\bibitem{Barnich:2000zw}
G.~Barnich, F.~Brandt, and M.~Henneaux, ``{Local BRST cohomology in gauge theories},'' {\em Phys. Rept.} {\bf 338} (2000) 439--569,
\href{http://arXiv.org/abs/hep-th/0002245}{{\tt hep-th/0002245}}.

\bibitem{Bergamin:2002ju}
L.~Bergamin and W.~Kummer, ``{Graded Poisson sigma models and dilaton-deformed 2d supergravity algebra},'' {\em JHEP} {\bf 05} (2003) 074,
\href{http://www.arXiv.org/abs/hep-th/0209209}{{\tt hep-th/0209209}}.

\bibitem{Hartong:2015xda}
J.~Hartong, ``{Gauging the Carroll Algebra and Ultra-Relativistic Gravity},'' {\em JHEP} {\bf 08} (2015) 069, \href{http://www.arXiv.org/abs/1505.05011}{{\tt 1505.05011}}.

\bibitem{Grumiller:2021cwg}
D.~Grumiller, R.~Ruzziconi, and C.~Zwikel, ``{Generalized dilaton gravity in 2d},'' {\em SciPost Phys.} {\bf 12} (2022), no.~1, 032, \href{http://www.arXiv.org/abs/2109.03266}{{\tt 2109.03266}}.

\bibitem{Bergamin:2003am}
L.~Bergamin and W.~Kummer, ``{The complete solution of 2D superfield supergravity from graded Poisson-Sigma models and the super pointparticle},'' {\em Phys. Rev.} {\bf D68} (2003) 104005,
\href{http://www.arXiv.org/abs/hep-th/0306217}{{\tt hep-th/0306217}}.

\bibitem{Coussaert:1995zp}
O.~Coussaert, M.~Henneaux, and P.~van Driel, ``{The Asymptotic dynamics of three-dimensional Einstein gravity with a negative cosmological constant},'' {\em Class.Quant.Grav.} {\bf 12} (1995) 2961--2966,
\href{http://www.arXiv.org/abs/gr-qc/9506019}{{\tt gr-qc/9506019}}.

\bibitem{Cadoni:1999ja}
M.~Cadoni and S.~Mignemi, ``{Asymptotic symmetries of AdS(2) and conformal group in d = 1},'' {\em Nucl. Phys.} {\bf B557} (1999) 165--180,
\href{http://www.arXiv.org/abs/hep-th/9902040}{{\tt hep-th/9902040}}.

\bibitem{Lodato:2016alv}
I.~Lodato and W.~Merbis, ``{Super-BMS$_{3}$ algebras from $ \mathcal{N}=2 $ flat supergravities},'' {\em JHEP} {\bf 11} (2016) 150, \href{http://www.arXiv.org/abs/1610.07506}{{\tt 1610.07506}}.

\bibitem{Bergamin:2004sr}
L.~Bergamin and W.~Kummer, ``Two-dimensional {N}=(2,2) dilaton supergravity from graded {P}oisson-{S}igma models {I}: Complete actions and their symmetries.,'' {\em Eur. Phys. J.} {\bf C39} (2005) S41--S52,
\href{http://www.arXiv.org/abs/hep-th/0402138}{{\tt hep-th/0402138}}.

\bibitem{Kitaev:15ur}
A.~Kitaev, ``{A simple model of quantum holography}.'' KITP strings seminars, April/May 2015, \href{http://online.kitp.ucsb.edu/online/entangled15/}{http://online.kitp.ucsb.edu/online/entangled15/} and \href{http://online.kitp.ucsb.edu/online/entangled15/kitaev2/}{http://online.kitp.ucsb.edu/online/entangled15/kitaev2/}.

\bibitem{Sachdev:1992fk}
S.~Sachdev and J.~Ye, ``{Gapless spin fluid ground state in a random, quantum Heisenberg magnet},'' {\em Phys. Rev. Lett.} {\bf 70} (1993) 3339,
\href{http://www.arXiv.org/abs/cond-mat/9212030}{{\tt cond-mat/9212030}}.

\bibitem{Sachdev:2010um}
S.~Sachdev, ``{Holographic metals and the fractionalized Fermi liquid},'' {\em Phys. Rev. Lett.} {\bf 105} (2010) 151602,
\href{http://www.arXiv.org/abs/1006.3794}{{\tt 1006.3794}}.

\bibitem{Maldacena:2016hyu}
J.~Maldacena and D.~Stanford, ``{Remarks on the Sachdev-Ye-Kitaev model},'' {\em Phys. Rev. D} {\bf 94} (2016), no.~10, 106002, \href{http://www.arXiv.org/abs/1604.07818}{{\tt 1604.07818}}.

\bibitem{Mertens:2018fds}
T.~G. Mertens, ``{The Schwarzian theory — origins},'' {\em JHEP} {\bf 05} (2018) 036,
\href{http://www.arXiv.org/abs/1801.09605}{{\tt 1801.09605}}.

\bibitem{Sarosi:2017ykf}
G.~S\'arosi, ``{AdS$_{2}$ holography and the SYK model},'' {\em PoS} {\bf Modave2017} (2018) 001,
\href{http://www.arXiv.org/abs/1711.08482}{{\tt 1711.08482}}.

\bibitem{Gu:2019jub}
Y.~Gu, A.~Kitaev, S.~Sachdev, and G.~Tarnopolsky, ``{Notes on the complex Sachdev-Ye-Kitaev model},'' {\em JHEP} {\bf 02} (2020) 157, \href{http://www.arXiv.org/abs/1910.14099}{{\tt 1910.14099}}.

\bibitem{Grumiller:2002nm}
D.~Grumiller, W.~Kummer, and D.~V. Vassilevich, ``Dilaton gravity in two dimensions,'' {\em Phys. Rept.} {\bf 369} (2002) 327--429,
\href{http://arXiv.org/abs/hep-th/0204253}{{\tt hep-th/0204253}}.

\bibitem{Pestun:2007rz}
V.~Pestun, ``{Localization of gauge theory on a four-sphere and supersymmetric Wilson loops},'' \href{http://www.arXiv.org/abs/0712.2824}{{\tt 0712.2824}}.

\bibitem{Witten:1981mf}
E.~Witten, ``A simple proof of the positive energy theorem,'' {\em Commun. Math. Phys.} {\bf 80} (1981)
381.

\bibitem{Grumiller:2017qao}
D.~Grumiller, R.~McNees, J.~Salzer, C.~Valc\'arcel, and D.~Vassilevich, ``{Menagerie of AdS$_{2}$ boundary conditions},'' {\em JHEP} {\bf 10} (2017) 203, \href{http://www.arXiv.org/abs/1708.08471}{{\tt 1708.08471}}.

\bibitem{Freedman:2012zz}
D.~Z. Freedman and A.~Van~Proeyen, {\em {Supergravity}}.
\newblock Cambridge Univ. Press, Cambridge, UK, 5, 2012.

\bibitem{Banerjee:2022ocj}
A.~Banerjee, S.~Dutta, and S.~Mondal, ``{Carroll fermions in two dimensions},'' {\em Phys. Rev. D} {\bf 107} (2023), no.~12, 125020, \href{http://www.arXiv.org/abs/2211.11639}{{\tt 2211.11639}}.

\bibitem{Bagchi:2022eui}
A.~Bagchi, A.~Banerjee, R.~Basu, M.~Islam, and S.~Mondal, ``{Magic fermions: Carroll and flat bands},'' {\em JHEP} {\bf 03} (2023) 227, \href{http://www.arXiv.org/abs/2211.11640}{{\tt 2211.11640}}.

\end{thebibliography}
\end{document}